\documentclass[
 onecolumn,
superscriptaddress,
 amsmath,amssymb,
 aps,
]{revtex4-2}

\usepackage{multirow}%
\usepackage[compact]{titlesec}
\titlespacing{\section}{0pt}{3ex}{3ex}
\titlespacing{\subsection}{0pt}{2ex}{2ex}
\titlespacing{\subsubsection}{0pt}{1ex}{1ex}
\usepackage{graphicx}
\usepackage{dcolumn}
\usepackage{bm}
\usepackage{booktabs, makecell, multirow}
\usepackage{booktabs,makecell,threeparttable}

\usepackage{subfigure}
\usepackage[dvipsnames]{xcolor}
\usepackage{comment}
\usepackage{xfrac}
\usepackage{amsmath, amssymb,bbding,oplotsymbl,MnSymbol}
\usepackage{lineno}

\usepackage{hyperref}
\hypersetup{
    colorlinks = true,
    urlcolor   = blue,
    citecolor  = red,
    linkcolor=   orange,
}

\DeclareSymbolFont{symbolsC}{U}{pxsyc}{m}{n}
\SetSymbolFont{symbolsC}{bold}{U}{pxsyc}{bx}{n}
\DeclareFontSubstitution{U}{pxsyc}{m}{n}
\DeclareMathSymbol{\medcirc}{\mathbin}{symbolsC}{7}
\DeclareFontSubstitution{OMS}{pxsy}{m}{n}

\definecolor{maroon}{rgb}{0.64, 0.08, 0.18}	

\definecolor{orange1}{rgb}{0.99, 0.50, 0.07}	
\definecolor{blue1}{rgb}{0.01, 0.21, 1.0}	
\definecolor{pink1}{rgb}{0.93, 0.01, 0.99}	
\definecolor{purple1}{rgb}{0.40, 0.02, 0.98}

\definecolor{d_cyan}{rgb}{0.0, 0.45,0.74}

\def\dpopa#1#2{{\frac{\partial#1}{\partial#2}}}
\def\spopa#1#2{{\frac{d#1}{d#2}}}

\def\spopasq#1#2{{{d^2#1}\over{d#2^2}}}

\newcommand{\bu}{\textbf{u}}

\newcommand{\bnabla}{\boldsymbol{\nabla}}

\newcommand{\RomanNumeralCaps}[1]

\begin{document}


\title{An Investigation of Flow and Interface Dynamics Near a Moving Contact Line at Obtuse Contact Angles}%

\author{Charul Gupta}
 \email{charul.gupta229@gmail.com}
\affiliation{Department of Mechanical \& Aerospace Engineering, Indian Institute of Technology Hyderabad, India\\
}%

\author{Venkata Sai Anvesh Sangadi}
\email{me20m22p100001@iith.ac.in}
\affiliation{Department of Mechanical \& Aerospace Engineering, Indian Institute of Technology Hyderabad, India\\
}%

\author{Lakshmana Dora Chandrala}
\email{lchandrala@mae.iith.ac.in}
\affiliation{Department of Mechanical \& Aerospace Engineering, Indian Institute of Technology Hyderabad, India\\
}%

\author{Harish N Dixit}
\email{hdixit@mae.iith.ac.in}
\affiliation{Department of Mechanical \& Aerospace Engineering, Indian Institute of Technology Hyderabad, India\\
}%
\affiliation{Center for Interdisciplinary Programs, Indian Institute of Technology Hyderabad, India\\
}%

\date{\today} 

\begin{abstract}
The flow near a moving contact line is primarily governed by three key parameters: viscosity ratio, dynamic contact angle, and inertia. While the behavior of dynamic contact angles has been extensively studied in earlier experimental and theoretical works, quantitative characterization of flow configurations remains limited. The present study reports detailed measurements of flow fields, interface shapes, and interfacial speeds in the low to moderate Reynolds number ($Re$) regimes using particle image velocimetry (PIV) and high-resolution image analysis. The investigation is restricted to dynamic contact angles greater than $90^{\circ}$. In the low-$Re$ regime, excellent agreement is observed between measured streamfunction contours and the modified viscous theory of Huh \& Scriven \cite{huh1971hydrodynamic} that accounts for a curved interface. Theoretical models such as the DRG formulation, using a single fitting parameter, accurately predict interface shapes even at finite $Re$. The interfacial speed away from the contact line compares favorably with theoretical predictions, whereas a pronounced deceleration is observed close to the contact line. Complementary Volume-of-Fluid (VoF) based numerical simulations were performed using identical geometric and material parameters to validate and extend the experimental observations. The simulations successfully reproduce the interface topology, flow structure, and the deceleration of the interfacial velocity near the contact line, providing strong support to the experimental findings. We argue that this rapid reduction in speed, observed both in experiments and simulations, is critical to the resolution of the long-standing moving contact line singularity. 
\begin{description}
\item[Keywords]
Stokes flow, dynamic contact angle, contact line dynamics.
\end{description}
\end{abstract}

\maketitle

\section{\label{sec:sec1}Introduction}
Moving contact line dynamics holds fundamental importance across various fluid flow scenarios. The basic mechanism underlying the dynamics involves the displacement of one fluid phase by another fluid phase (phases A and B) over a solid surface. Numerous applications, ranging from natural phenomena like drops traversing lotus leaves or spreading across solid surfaces to industrial processes such as the deposition of tiny drops onto a piece of paper in inkjet printing, exhibit a similar mechanism. All of these scenarios share a common feature: a moving contact line. The classical work of Huh \& Scriven\cite{huh1971hydrodynamic} suggested that only two parameters are sufficient to determine the flow field in the vicinity of a moving contact line, the dynamic contact angle ($\theta_d$) and viscosity ratio ($\mu_A/\mu_B$). But this theoretical model suffers from a singularity at the contact line. Several theoretical models have been developed to relieve the singularity and also describe the relationship between the dynamic contact angle and the speed of the contact line \cite{blake1969kinetics,de1985wetting,voinov1976hydrodynamics,cox1986dynamics,shikhmurzaev1993moving}. For example, the viscous study of Cox \cite{cox1986dynamics} associated the dynamic contact angle with the capillary number and the logarithm of the ratio of length scales, i.e., $Ca \ln({l_{\text{macro}}/l_s})$, where $l_{\text{macro}}$ and $l_s$ are the macroscopic and microscopic length scales, respectively. Several other models have also been proposed over the years and these are summarised in excellent reviews and monographs \cite{dussan1979spreading,snoeijer2013moving,shikhmurzaev2007capillary}.

Few theoretical studies have focused on the flow patterns emerging near the moving contact line. These studies were conducted in the viscous regime by ensuring that both the Reynolds number ($Re= \rho_B U l_{\text{macro}}/\mu_B$) and the capillary number ($Ca= \mu_B U/\sigma$) are much less than unity, i.e., $Re \ll 1$ and $Ca \ll 1$. Here, the subscript B denotes fluid phase B, while $\rho$ represents density, $\mu$ denotes viscosity, $\sigma$ stands for surface tension, and $U$ refers to plate speed. The study by Huh \& Scriven (HS71 hereafter) \cite{huh1971hydrodynamic} predicted the flow configurations in a local region near the moving contact line by solving the biharmonic equation, assuming standard no-slip boundary condition at the moving solid. One of the key predictions of the HS71 theory was the occurrence of a rolling-type motion in the more viscous phase. However, the theory also suffers a singularity at the moving contact line, where the shear stress diverges. Subsequent theoretical studies were conducted, aiming to circumvent the singularity by incorporating new physics at the contact line such as allowing the contact line to slip over the solid or employing a precursor thin film over the solid. For example, the study by Cox \cite{cox1986dynamics} employed a constant slip in the slip region near the contact line. Cox divided the local region into three smaller regions using the method of matched asymptotic expansion: a slip, intermediate and outer regions and predicted flow patterns in the intermediate region. The predictions of flow configurations from Cox and other similar theories \cite{shikhmurzaev1997moving} align closely with the predictions of the HS71 theory. This underscores the significance of HS71 theory and serves as a motivation to test it. The successful testing of the theory will also validate other theoretical studies.

Recent theoretical studies by Kirkinis and Davis \cite{kirkinis2013hydrodynamic,kirkinis2014moffatt} employed a variable-slip model over a finite distance from the contact line while maintaining a no-slip condition over the remaining solid surface. Unlike Kirkinis \textit{et al.}, who examined single-phase flow, Febres \textit{et al.} \cite{febres2017existence} extended the problem to two-phase flows, predicting inner-region flow configurations by solving an eigenvalue problem analogous to that of Moffatt \cite{moffatt1964viscous}. Their solution yielded multiple admissible values of $n$, each corresponding to a distinct flow configuration, thereby complicating model selection. Dynamic interface shapes near the contact line have also been addressed in the literature using theoretical models. Dussan \textit{et al.} \cite{rame1991identifying} presented a composite formulation combining Cox theory \cite{cox1986dynamics} with the static interface shape to predict profiles across a range of length scales. In contrast, Chan \textit{et al.} \cite{chan2013hydrodynamics,chan2020cox} solved a coupled system of differential equations with slip along the solid surface, applying boundary conditions that enforced an equilibrium contact angle at the contact line and a flat interface in the far field. Both approaches required empirical tuning parameters, limiting their predictive capability. More recently, Kulkarni \textit{et al.} \cite{kulkarni2023stream} obtained streamfunction solutions under various slip boundary conditions. A detailed discussion of selected theoretical models is provided in \S\ref{sec:theory_background}.

Earlier theoretical efforts have addressed diverse aspects of moving contact line dynamics, including dynamic contact angles, flow fields, interface shapes, and interfacial velocities, with complementary experimental studies exploring similar phenomena. Dussan \textit{et al.} \cite{dussan1974motion} inferred global flow structures within liquid drops (honey and glycerol) moving over an inclined plane by tracking interface-adjacent tracer particles. The classical work of Hoffman \cite{hoffman1975study} demonstrated the universal dependence of dynamic contact angle on capillary number by driving immiscible fluid systems through capillaries. Several investigations \cite{rame1991identifying, chen1995breakdown, rame1996microscopic}, including those by Dussan \textit{et al.}, revealed deviations of dynamic interface shapes from static configurations due to viscous effects near the contact line. Chen \textit{et al.} \cite{chen1997velocity} reported PIV measurements using a 60,000 cSt PDMS liquid in the viscous regime, identifying a \textit{rolling} flow structure near the contact line. Other studies \cite{le2005shape,rio2005boundary} measured dynamic contact angles during drop sliding and compared them against various viscous-regime models, whereas Puthenveettil \textit{et al.} \cite{puthenveettil2013motion} conducted similar experiments with mercury and water drops in the inertial regime. Gupta \textit{et al.} \cite{gupta2022study} observed negligible qualitative influence of $Re$ on flow configuration in a limited set of experiments with dynamic contact angles below $90^{\circ}$. Their subsequent work \cite{gupta2023} performed detailed viscous-regime measurements at low viscosity ratio ($\lambda \ll 1$) and acute contact angles, while Gupta \textit{et al.} \cite{gupta2024experimental} conducted the first two-phase PIV measurements for $\lambda > 1$, focusing on flow configurations. Both studies jointly measured flow fields, interface shapes, and interfacial velocities, providing rigorous tests of theoretical models. The present study extends these efforts to the obtuse-angle regime, with dynamic contact angles strictly above $90^{\circ}$, low viscosity ratio, and variable $Re$ and capillary numbers achieved by adjusting the plate velocity. In addition to experiments, numerical simulations are employed to confirm that the physical mechanisms observed experimentally can be reproduced computationally.

The framework employed in the simulations closely follows the work of Fullana \emph{et al.}\cite{fullana2024consistent}, but we restrict the simulations to the simpler Navier-slip model with a spatially varying slip coefficient. While more sophisticated models have also been developed to address the problem of singularity at the contact line in a self-consistent way, we find that the Navier-slip model adequately captures the flow fields observed in experiments.
These simulations, configured to replicate the experimental geometry and boundary conditions, allow direct comparison of predicted and measured flow fields, interface shapes, and interfacial speeds in the obtuse-angle regime, thereby reinforcing the consistency of the observed dynamics across both approaches.

In this work, we systematically investigate flow dynamics by varying viscosity ratio and dynamic contact angle. All experiments are conducted with dynamic contact angles above $90^{\circ}$ and viscosity ratios below unity. A vertically translating plate is immersed in a liquid bath at controlled speeds to achieve low to moderate $Re$. To the best of our knowledge, no comparable experiments have been reported, apart from earlier studies \cite{dussan1974motion,chen1997velocity} using highly viscous liquids at very low speeds in the viscous regime. In parallel, numerical simulations are performed with the same geometry and boundary conditions as the experiments, serving to verify that the computational model reproduces the key physical features of the observed dynamics. This combined approach enables us to probe the influence of $Re$ on moving contact line behaviour using a plate-immersion configuration that permits high-fidelity 2D PIV measurements, which are impractical in other setups due to optical limitations. We quantify flow fields, interface shapes, and interfacial velocities near the contact line, comparing them rigorously with theoretical predictions. While our methodology follows Gupta \textit{et al.} \cite{gupta2023} for acute contact angles, the motivation for a dedicated obtuse-angle study is twofold: (i) to explore potentially distinct dynamics, as suggested by Duez \textit{et al.} \cite{duez2007making}, and (ii) because water and aqueous systems—common in nature—tend to form obtuse dynamic contact angles during motion on solid surfaces. This study provides, for the first time, a systematic assessment of $Re$ effects on moving contact line behaviour across the parameter space in Table \ref{tab:operating_parameters}. The measured interfacial velocities and interface shapes near the contact line offer boundary condition data for future theoretical and numerical modelling.

The paper is structured as follows: \S\ref{sec:experimental_setup} outlines the experimental methodology and the cleaning protocol. A discussion on the theories related to flow fields and interface shape models is provided in \S\ref{sec:theory_background}. \S\ref{sec:numerics} describes the numerical methodology, included primarily to complement the experiments and verify that the observed dynamics are captured computationally. In \S\ref{sec:results}, comparisons of flow fields, interface shapes, and interfacial speeds against existing theoretical models are presented, including corresponding results from the simulations. Finally, the key findings and important observations are discussed in \S\ref{sec:conclusion}.

\section{Experimental Methodology}
\label{sec:experimental_setup}
\begin{figure}
\centering
\includegraphics[trim = 0mm 0mm 0mm 0mm, clip, angle=0,width=0.6\textwidth]{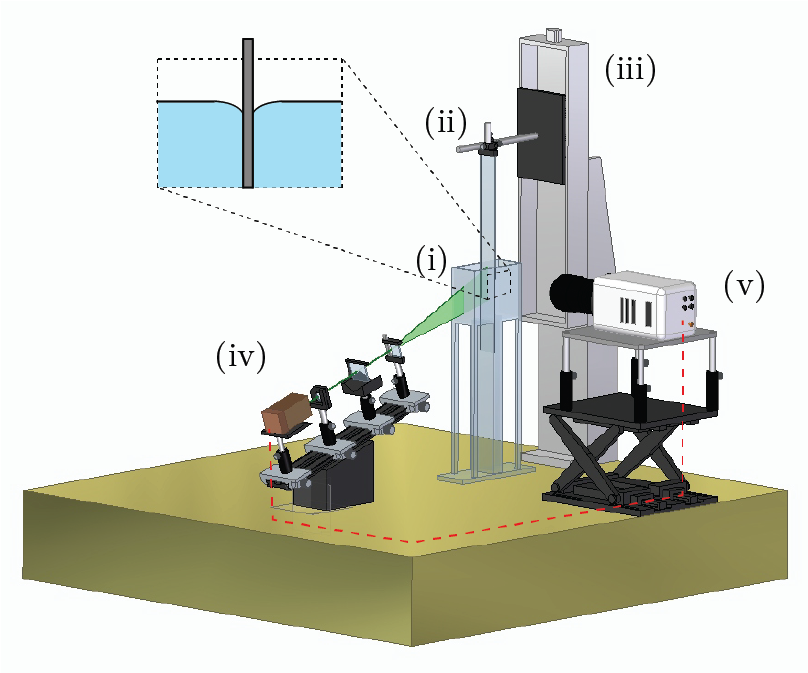}
\caption{ An illustration of the experimental setup. (i) A transparent acrylic tank filled with liquid (ii) A glass plate traversing into the liquid bath (iii) A linear traversing system (iv) A combination of laser and lens for forming a thin laser sheet (v) High-speed camera. The inset provides a magnified view of the field of interest, showing the fluid phase B forming an obtuse contact angle with the solid surface. }
\label{fig:experimental_setup}
\end{figure} 
 The experimental setup is illustrated in figure \ref{fig:experimental_setup}. To facilitate optical measurements, experiments were conducted within a transparent rectangular acrylic tank measuring $100 \times 100 \times 26$ mm³. The tank was filled with liquid (phase B) to a depth of 80 mm. Various liquids, including silicone oil, water, and sugar-water mixtures, were employed to cover a wide range of viscosity ratios, with their properties summarized in Table \ref{tab:properties}. Before the start of each experiment, a glass slide measuring $75\text{mm}\times25\text{mm}\times1\text{mm}$ was partially submerged in the liquid pool, establishing a contact line at the intersection of the two phases with the solid substrate. The immersion speed of the solid substrate/plate was controlled by connecting it to a programmable linear traverse. This enabled precise variation of the plate speed from a few microns per second to several centimetres per second, thereby generating a wide range of Reynolds numbers.  In order to keep the liquid level constant as the plate submerged, a syringe pump (not shown in figure \ref{fig:experimental_setup}) was used to extract an equivalent volume from the tank.

\begin{table}[h]
\centering 
\caption{Operating parameters}
\begin{tabular}{ c c  c  c c c c} 
\toprule
      Fluid-fluid system &  
      \begin{tabular}{@{}c@{}}Plate speed \\ $U$(mm/s)
      \end{tabular}  &
      \begin{tabular}{@{}c@{}}Reynolds number\\ ($Re$) \end{tabular}  &
      \begin{tabular}{@{}c@{}}Capillary number \\ ($Ca$)  \end{tabular}  &
       \begin{tabular}{@{}c@{}}Dynamic contact angle \\ $\theta_d$, (deg)\end{tabular}  & 
       \begin{tabular}{@{}c@{}}Static advancing angle \\ $\theta_{sa}$, (deg)\end{tabular}  & 
      \begin{tabular}{@{}c@{}} Surface \\ property \\ \end{tabular}  \\
 \midrule
    \multirow{4}{*}{Air--water}
        & 0.5 & 1.52 & $6.18\times10^{-6}$ & $\approx94.2$ & $\approx94$ & Coated \\
        & 1.0 & 3.03 & $1.24\times10^{-5}$ & $\approx97.7$ & -- & Coated \\
        & 2.0 & 6.07 & $2.47\times10^{-5}$ & $\approx102.6$ & -- & Coated \\
        & 5.0 & 15.2 & $6.18\times10^{-5}$ & $\approx103.6$ & -- & Coated \\
    \midrule
    \multirow{3}{*}{Air--48\% sugar--water}
        & 0.5 & $1.25\times10^{-1}$ & $8.16\times10^{-5}$ & $\approx95.5$ & $\approx95$ & Coated \\
        & 1.0 & $2.50\times10^{-1}$ & $1.63\times10^{-4}$ & $\approx95.7$ & -- & Coated \\
        & 2.0 & $5.00\times10^{-1}$ & $3.26\times10^{-4}$ & $\approx101.1$ & -- & Coated \\
    \midrule
    \multirow{2}{*}{Air--60\% sugar--water}
        & 0.10 & $5.95\times10^{-3}$ & $6.96\times10^{-5}$ & $\approx99.1$ & $\approx99$ & Coated \\
        & 0.15 & $8.92\times10^{-3}$ & $1.04\times10^{-4}$ & $\approx105.2$ & -- & Coated \\
    \midrule
    \multirow{2}{*}{Air--500\,cSt silicone oil}
        & 2.0 & $5.23\times10^{-3}$ & $5.58\times10^{-2}$ & $\approx96.8$ & $\approx12$ & Uncoated \\
        & 3.0 & $7.85\times10^{-3}$ & $8.38\times10^{-2}$ & $\approx110.4$ & -- & Uncoated \\
    \bottomrule
\label{tab:operating_parameters}
\end{tabular}
\end{table}
 
\begin{table}[h]
\centering 
\caption{Fluid properties.}
\begin{tabular}{ c @{\hskip 0.05in} c  c  c  c } 
\hline
      &  \begin{tabular}{@{}c@{}}Density \\ $kg/m^3$\end{tabular} & 
      \begin{tabular}{@{}c@{}}Viscosity \\ $10^{-3} Pa.s$\end{tabular}  & 
      \begin{tabular}{@{}c@{}}Surface tension \\ mN.m\end{tabular}  \\
\hline
Air & 1.207 & 0.0189 & - \\ 
Water & 1000 & 0.89 & 72 \\ 
48$\%$ (w/w) Sugar water mixture & 1241.76 & 12.31 & 75.42 \\
60$\%$ (w/w) Sugar water mixture & 1330 & 54.77 & 78.71 \\
500 cSt Silicone oil & 965 & 516.67 & 18.5\\
\hline
\label{tab:properties}
\end{tabular}
\end{table}

%
Water and water-based liquids typically form an acute static contact angle over the glass slide. In experiments involving water-based liquids, a hydrophobic coating was applied to the slides to achieve an obtuse contact angle. However, in silicone oil experiments,  the obtuse contact angle was achieved by increasing the speed of the slide without requiring any coating on the glass slide. The surface roughness of the slides was determined using a profilometer and found to be in the range of 20 nanometers.

%
\begin{figure}[h]
\centering
\includegraphics[trim = 0mm 0mm 0mm 0mm, clip, angle=0,width=0.4\textwidth]{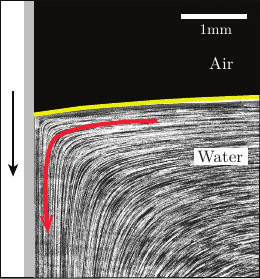}
\vspace{2mm}
\caption{A streakline image illustrates the flow pattern in the vicinity of a moving contact line at $Re= 3.03$ and $Ca = 1.24\times 10^{-5}$. As the solid surface, depicted by a grey slab, descends into the liquid, the material points at the interface migrate toward the contact line and subsequently align with the solid surface. The red arrow indicates the flow direction, and the yellow solid curve indicates the air and water interface.}
\label{fig:Water_streakline_image}
\end{figure}

In all the experiments, the same cleaning protocol was followed to eliminate variations. At first, the tank and glass slides were rinsed with distilled water thoroughly and dried using a dryer to avoid any liquid residue. Subsequently, the plate was coated with a hydrophobic coating to obtain a high contact angle over the plate. Following the coating process, the plates were once again rinsed with distilled water and dried with a dryer, which allowed sufficient time for the plate to be in equilibrium with room temperature.

A two-dimensional particle image velocimetry (PIV) technique was utilized to measure the velocity field near the contact line in the mid-plane of the tank (see figure \ref{fig:experimental_setup}). The PIV system comprises a Photron Nova S9 high-speed camera with a resolution of 1024 pixels × 1024 pixels attached with a macro lens, a green continuous wave diode laser with a maximum power of 2W, and sheet optics. The sheet optics,  a combination of biconcave spherical and plano-concave cylindrical lenses, produces a laser sheet of about 0.5 mm thick. In water and water-based experiments, polystyrene particles with an average diameter of 2$\mu$m were used as tracers, while for silicone oil experiments, polyamide particles with an average diameter of 5$\mu$m were used. The acquisition frame rate of the high-speed camera varied between 50 and 1000 frames per second, depending on the plate speed. 

The high-speed camera, coupled with a macro lens attachment, captured a field of view measuring approximately $4.2\text{mm}\times 4.2\text{mm}$, yielding a spatial resolution of $4\mu$m per pixel.  Several image processing steps were performed on raw PIV images including average background subtraction and image equalization before a cross-correlation analysis was carried out. Since the flow is steady, the pre-processed PIV images were analyzed using a multigrid and window-deforming ensemble correlation method. A total of 500 images were used to perform ensemble correlation with a final interrogation window size of 16 pixels x 16 pixels.

A sample  streaklines image at $Re= 3.03$ and $Ca = 1.24\times 10^{-5}$ is shown in figure~\ref{fig:Water_streakline_image}. This image was generated by overlaying 500 consecutive particle images and averaging their maximum intensities. The absence of intersecting streaklines indicates a steady flow. The streakline image is further utilized to extract the interface shape and the extracted points were fitted with a two-term exponential function given by
\begin{equation}\label{eq:fitting_eq}
   y = ae^{bx} + ce^{dx} 
\end{equation}
Here $y$ is the vertical location of the interface (shown with a yellow curve in figure \ref{fig:Water_streakline_image}) and $x$ is the horizontal distance measured from the vertical plate. The constants $a$, $b$, $c$, and $d$ were determined for each experiment, and in all cases, the two-term exponential fit resulted in an R-square value of close to 0.99. The main purpose of fitting such a function \eqref{eq:fitting_eq} was to extract the local interface angle (slope) which is used in various dynamic contact angle models.

\section{\label{sec:theory_background}Theoretical background}
\begin{figure}
\centering
\subfigure[]
{\label{fig:geometry_huh}
\includegraphics[trim = 0mm 0mm 0mm 0mm, clip, angle=0,width=0.27\textwidth]{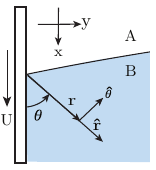}}
\hspace{20mm}
\subfigure[]
{\label{fig:rolling_flow_schematic}
\includegraphics[trim = 0mm 0mm 0mm 0mm, clip, angle=0,width=0.22\textwidth]{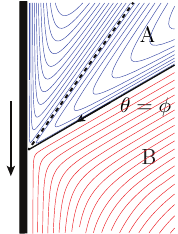}}
\caption{A schematic diagram of the problem geometry and the flow patterns emerged near the contact line. (a) A solid surface, represented by a black-edged slab, descends into the liquid bath in phase B. The system is characterized by $\hat{r}-\hat{\theta}$ cylindrical coordinate system, where $\hat{r}$ and $\hat{\theta}$ represent the radial and angular coordinates, respectively, with the contact point serving as the origin. An interface between phase A and phase B is depicted by a solid black line at $\theta=\phi$. (b) Considering the advancing motion of the solid, kinematically consistent flow patterns are shown in both phases: split-streamline motion in phase A and rolling motion in phase B.}
\label{fig:schematic_analytical}
\end{figure}

Several theoretical studies have been carried out in the local region near the contact line. Figure \ref{fig:geometry_huh} illustrates this local region at a length scale $L$, such that $Re_{L} \ll 1$ and $Ca \ll 1$. The study by HS71 solved the biharmonic equation in the same local region and revealed two key parameters: viscosity ratio and dynamic contact angle. They reported several flow configurations in different regimes of the parameter spaces. A typical flow configuration at an obtuse dynamic contact angle and low viscosity ratio is shown in figure \ref{fig:rolling_flow_schematic}. Here, fluid particles at the interface move toward the contact line, resulting in a rolling motion in phase B and a split-streamline motion in phase A. Even though HS71 reported stable flow patterns near the contact line, the theory suffers from a stress singularity at the contact line, i.e., $\tau|_{r\rightarrow 0,\theta = 0}\rightarrow \infty$ where $\tau$ represents the shear stress. The angle $\theta$ is measured from the solid surface in the liquid phase B with the contact line as the origin. Subsequent theoretical studies were conducted using different approaches concentrating on relieving the singularity at the contact line. Several approaches, such as employing a constant slip or a precursor film over the solid near the contact line were utilized. The seminal study of Cox \cite{cox1986dynamics} employed a constant slip by dividing the region near the contact line into three sub-regions as shown in figure \ref{fig:split_regions}, and the solution was obtained using the method of matched asymptotic expansion. The inner region is dominated by slip at the contact line while the outer region depends on the geometry of the problem. The intermediate region, where the no-slip boundary condition applies, matches the inner and outer regions as shown in figure  \ref{fig:split_regions}.

Based on the study by Dussan \textit{et al.} \cite{rame1991identifying}, the inner region near the moving contact line governs the flow dynamics in the intermediate and outer regions. However, the reverse is not true. Since the inner region is beyond the range of optical imaging capabilities, understanding the flow configuration in the intermediate region can assist in inferring the configuration in the inner region as well. To ensure the existence of the intermediate region between the inner and outer regions, the following expression, $|Ca \ln (l_s/l_{\text{macro}})| \ll 2$, should hold as explicitly mentioned in the study of Sibley \textit{et al.} \cite{sibley2015asymptotics}. The length scale for the intermediate region ($l_i$) should satisfy $l_s \ll l_i < l_{\text{macro}}$, where $l_s$ denotes the slip length. In all our experiments, the capillary number, $Ca$, is kept low ($Ca<10^{-2}$), and slip length typically falls within the nanometer range. Therefore, the above condition for the presence of a wide intermediate region would always be satisfied and is accessible in our experiments. The shape of the interface and flow field in the intermediate region can thus can be compared against theoretical predictions. 

Below we outline the models for interface shape and flow field, and this forms the basis for comparison between experiments and theoretical predictions.

\subsection{Theoretical model for the interface shape}\label{sec:Interface_Shape_theory} 
In a static condition, the interface deforms near the contact line due to balance of capillary and gravity forces. However, in a dynamic condition, viscous forces act primarily near the contact line and deforms the interface from its static shape. The deviation of the dynamic interface shape from the static shape is denoted as viscous deformation and is confined to a small region near the moving contact line. Away from the contact line, a balance between surface tension and gravity emerges and the interface shape resembles that of a static meniscus. In the present experimental configuration shown in figure \ref{fig:experimental_setup}, the interface becomes flat and horizontal far away from the moving plate. A model that aims to capture the interface shape in a dynamic setting is required to capture physics both at the moving plate as well as in the far field. Below, we will review two popular models that capture interface shape.
%
\begin{figure}[h]
\centering
\subfigure[]
{\label{fig:split_regions}
  \includegraphics[trim = 0mm 0mm 0mm 0mm, clip, angle=0,width=0.22\textwidth]{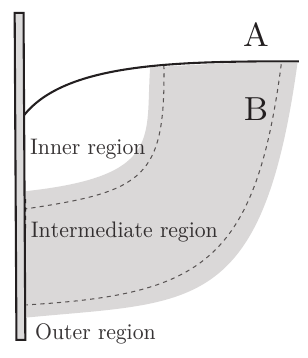}
}
\hspace{20mm}
\subfigure[]
{\label{fig:coordinate_system}
  \includegraphics[trim = 0mm 0mm 0mm 0mm, clip, angle=0,width=0.28\textwidth]{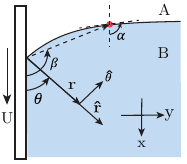}
}
\caption{(a) The cartoon depicts three regions as proposed by the study of Cox \cite{cox1986dynamics}. (b) The schematic indicates the problem geometry with the curved interface where the angle $\beta$ represents the interface angle shown for an arbitrary point on the interface. The angle $\alpha$ represents a local interface slope measured from the vertical direction.}
\end{figure}

\begin{figure}[h]
\centering
\subfigure[]{
\label{fig:omega_o}
\includegraphics[trim = 0mm 0mm 0mm 0mm, clip, angle=0,width=0.31\textwidth]{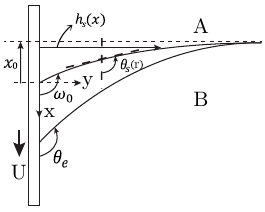}
}
\hspace{15mm}
\subfigure[]{
\label{fig:omega_o_GLM}
\includegraphics[trim = 0mm 0mm 0mm 0mm, clip, angle=0,width=0.31\textwidth]{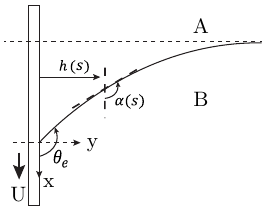}
}
\caption{(a) A schematic represents two interfaces: one with the equilibrium contact angle $\theta_e$ at the wall and the other with $\omega_0$ angle at the wall. The latter interface represents the static interface dictated by the outer solution $h_s(x)$ with the contact point as the origin at the wall. The contact angle $\omega_0$ is obtained by matching the inner to the outer region as mentioned in \cite{rame1991identifying}. (b) depicts an interface that forms an equilibrium angle ($\theta_e$) at the contact line while becoming flat in the far field. The interface is dictated by $h(s)$ at any arbitrary arc length $s$ from the origin where contact line is at origin. $\alpha(s)$ denotes a local slope at that arbitrary arc length.} 
\end{figure}
%

%
Dussan \emph{et al.}\cite{rame1991identifying} proposed a composite solution for dynamic interface shape by integrating the Cox model, figure \ref{fig:split_regions}, with the static interface shape. The Cox model accounts for viscous effects, whereas the static shape is influenced by capillary and gravity forces. The incorporation of the static shape enables this model, referred to as DRG model, to predict the interface shape at various length scales. The model predicts the interface shape in terms of the local angle at the interface, i.e. $\alpha(r)$ shown in figure \ref{fig:coordinate_system}, at varying radial locations along the interface in the region spanning from the intermediate to the outer region. The DRG model can be expressed as follows:
\begin{equation}\label{eq:drgcomp}
   \alpha(r) = J^{-1}\left[J(\omega_0)+Ca\ln\left(\frac{r}{l_c}\right)\right] + \left(\theta_s - \omega_0\right),
\end{equation}
where $l_c$ denotes the characteristic length, i.e., capillary length and $\omega_0$ represents a tuning parameter. In \cite{rame1991identifying}, the value of $\omega_0$ was estimated by minimizing the difference between the interface shape obtained from experiments and the DRG model, i.e. equn. \eqref{eq:drgcomp}. Physically, one can view $\omega_0$ as the apparent or dynamic contact angle of the system when DRG model is in perfect agreement with experiments.

If viscosity ratio, $\lambda \rightarrow 0$, then the function $J(x)$ can be defined as
\begin{eqnarray}\label{eq:gtheta}
  J(x) = \int^x_0 \frac{p - \cos p \sin p}{2\sin p} dp,
\end{eqnarray}
The effect of $\lambda$ is assumed to be negligible for a liquid-gas interface. In equn. \eqref{eq:drgcomp}, the static interface shape, $\theta_s$, can be expressed by
 \begin{equation}\label{eq:outer}
    \theta_{s} = g_0\left(\frac{r}{l_c};\omega_0\right).
 \end{equation}
where $\theta_{s}$ is the local slope from the vertical measured along the static interface (see figure \ref{fig:omega_o}). The exact expression for the static shape, given by the function $g_0$, formed over flat plate geometry is given in \S \ref{appx:static_shape}. A detailed discussion on the comparison of interface shapes is provided in \S \ref{sec:interface_shape}.  

\begin{figure}[h]
\centering
\includegraphics[trim = 0mm 0mm 0mm 0mm, clip, angle=0,width=0.6\textwidth]{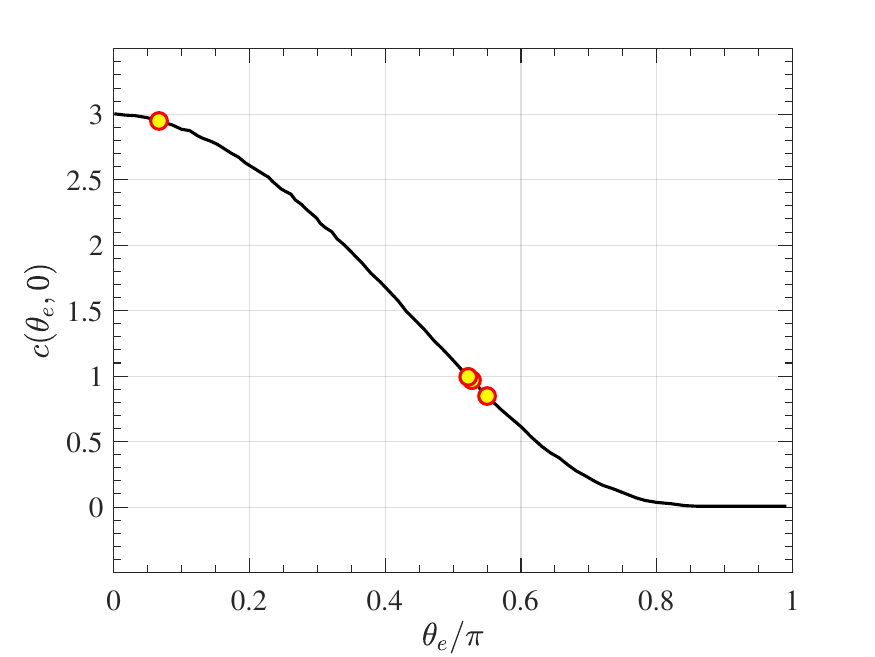}
\vspace{2mm}
\caption{Variation of c with the static advancing equilibrium angle. A solid black curve is reproduced using Chan \textit{et al.} \cite{chan2020cox} analysis and filled circles denote our data points where interface shape are compared with GLM as shown in figure \ref{fig:interface_shape_GLM}.}
\label{fig:c_vs_theta}
\end{figure}

In a recent development, Snoeijer and coworkers \citep{snoeijer2006free, chan2013hydrodynamics, chan2020cox} developed a generalized lubrication model (GLM) to determine the full interface shape. This model relies on a low capillary number expansion of the Stokes' equations around a constant wedge angle. Unlike classical lubrication theory which is restricted to small interface slopes, the GLM approach can be used for all interface angles. The model reduces to two coupled differential equations for the interface shape, $h(s)$, and the local interface angle, $\alpha(s)$, shown schematically in figure \ref{fig:omega_o_GLM},
\begin{subequations}
  \begin{align}
    \spopasq{\alpha}{s} = &~\frac{3 Ca}{h(h+cl_s)} f(\alpha,\lambda) - \frac{1}{l_c^2}\cos(\alpha),\\
    \spopa{h}{s} = &~\sin(\alpha),
 \end{align}
 \label{eq:glm}
\end{subequations}
where, $s$, is the arc length measured along the interface. The singularity at the contact line is regularized by imposing a slip, with slip-length $l_s$, at the contact line. The constant, $c$, depends on the equilibrium contact angle. Chan \emph{et al.} \cite{chan2020cox} showed that the constant $c$ is approximately equal to 3 for small contact angles and decreases to zero for large contact angles as shown in figure \ref{fig:c_vs_theta}.
While the general form of the function $f(\alpha,\lambda)$ can be easily used in the solution of equn. \eqref{eq:glm}, in the present study, $\lambda \ll 1$. We therefore use the $\lambda=0$ limit in the analysis and the function $f(\alpha,\lambda)$ simplifies to
\begin{equation}
    f(\alpha,0)=-\frac{2\sin^3(\alpha)}{3(\alpha-\sin(\alpha)\cos(\alpha))}.
\end{equation}
Using a small capillary number ($Ca \ll 1$), Snoeijer \textit{et al.} \cite{snoeijer2006free} integrated equn. \eqref{eq:glm} and recovered the expression of the interface shape reported by Dussan \textit{et al.} \cite{rame1991identifying}. While the DRG model requires matching the solutions from two distinct regions to determine the free parameter $\omega_0$ and requires input from the experimental interface shape, the GLM approach eliminates the need for such matching and is thus potentially more attractive. To solve \eqref{eq:glm}, the following boundary conditions on the moving plate and the far-field are employed:
\begin{eqnarray}
    \alpha(s=0)=\theta_e;\quad \alpha(s\to\infty)=\frac{\pi}{2}, \qquad h(s=0)=0. \label{eq:glmbc}
\end{eqnarray}
Both DRG model and GLM are compared in detail with experimental interface shapes in \S \ref{sec:interface_shape}.

\subsection{Theoretical models for flow fields} \label{sec:models_flow_fields}
The HS71 theory provided a solution in a local region near the moving contact line assuming the interface to be flat, with angle $\phi$, as shown in figure \ref{fig:geometry_huh}. The governing equation, written in the $r-\theta$ coordinate system is given by
\begin{equation}
    \bnabla^4 \psi = 0,
\end{equation}
where
\begin{equation}\label{eq:Scriven1}
   \psi(r,\theta;\phi) = r(a_1(\phi) \sin \theta + a_2(\phi) \cos \theta + a_3(\phi) \theta \sin \theta + a_4(\phi) \theta \cos \theta).
\end{equation}
Here, the coefficients $a_1,a_2,a_3 \,\,\,  \text{and} \, \,\, a_4$ are functions of the fixed wedge angle $\phi$. But in real flows, the interface is often curved. To facilitate a comparison between the HS71 theory and experiments, it is necessary to modify the above solution to accommodate a curved interface. Chen et al. \cite{chen1997velocity} proposed modifying the wedge angle, $\phi$, in equn. \eqref{eq:Scriven1} with the local interface angle, $\beta(r)$ shown schematically in figure \ref{fig:coordinate_system}. In a recent study \cite{gupta2023}, it was shown that this small change in the theoretical expression for streamfunction accounts for curvature effects that are absent in HS71 theory. The modified expression for streamfunction with a curved interface now becomes
\begin{equation}\label{eq:MWS_psi}
   \psi(r,\theta;\beta) = r(a_1(\beta) \sin \theta + a_2(\beta) \cos \theta + a_3(\beta) \theta \sin \theta + a_4(\beta) \theta \cos \theta).
\end{equation}
The expression for the coefficients $a_1(\beta) - a_4(\beta)$, is given in \S \ref{appx:fixed_wedge}. If the gas above the interface is assumed to be passive, i.e. viscosity ratio $\lambda \rightarrow 0$, then equn. \eqref{eq:MWS_psi} greatly simplifies to the form
%
\begin{equation}\label{eq:Modulate_wedge_1}
   \psi(r,\theta;\beta(r)) = rK(\theta,\beta) = rU\left(\frac{\theta\sin\beta \cos(\theta -\beta)-\beta\sin\theta}{\sin\beta \cos\beta - \beta}\right).
\end{equation}
%
The expression for streamfunction for a fixed wedge can simply be obtained by replacing $\beta(r)$ with the fixed wedge angle $\phi$ as given in equn. \eqref{eq:streamfunction_HS71} of \S \ref{appx:fixed_wedge}. The above expression for a curved interface was referred to as `modulated wedge solution' (MWS) by Chen \textit{et al.} \cite{chen1997velocity}. It has to be noted that the interface angle $\beta(r)$ approaches the fixed wedge angle $\phi$ at the contact line, i.e. $\beta(r)|_{r\rightarrow 0} \rightarrow \phi$. Therefore, the streamfunction expressions for HS71 and MWS theory become identical as one approaches the contact line which is consistent with the view that curvature effects become negligible in the vicinity of the contact line. Unlike in the case of a flat wedge where the interfacial speed coincides with the radial velocity (represented as $v_i^{HS}$ given by equn. \eqref{eq:Scriven2} in \S \ref{appx:fixed_wedge}), the interfacial speed along a curved interface will have both radial ($v_r$) and angular ($v_{\theta}$) velocity components (see \S \ref{appx:MWS} for the expressions) and can be written as
%
\begin{equation}\label{eq:modulated_interfacial_speed}
    v_i^{MWS} = v_r(r,\beta) \cos(\beta - \alpha) - v_{\theta}(r,\beta) \sin(\beta - \alpha),
\end{equation}
%
where $\alpha(r)$ represents the local slope of the interface from vertical (see figure \ref{fig:coordinate_system}) at a radial location $r$, related to $\beta(r)$ by the following expression:
\begin{equation} \label{eq:local_slope}
    \alpha(r) = \beta(r) + \tan^{-1}\left(r\spopa{\beta}{r}\right).
\end{equation}
Using simple geometrical arguments, the expression for interfacial speed can be simplified to 
\begin{equation} \label{eq:interfacial_speed_relation}
    v_i^{MWS} = \frac{U}{\cos( \beta- \alpha)}\left(\frac{\sin \beta-\beta \cos \beta}{\sin \beta \cos \beta - \beta}\right).
\end{equation}
%
Since the local interface angle $\beta(r)$ varies along the interface, the interfacial speed for a curved interface is not constant unlike the constant value found for a flat interface. Further, at the contact line, $\alpha \equiv \beta$. Thus the interfacial speed for a curved interface ($v_{i}^{MWS}$) becomes equal to that of a flat interface ($v_{i}^{HS}$) as $r\rightarrow 0$. 

Both streamfunctions and interfacial speeds are compared in detail with experimental data in \S \ref{sec:flow_fields_ex_MWS} and \S \ref{sec:interfacial_speed}, respectively.

\section{Numerical method} \label{sec:numerics}
\begin{figure}[h]
    \centering
        \includegraphics[trim=0mm 0mm 0mm 0mm, clip, width=0.65\textwidth]{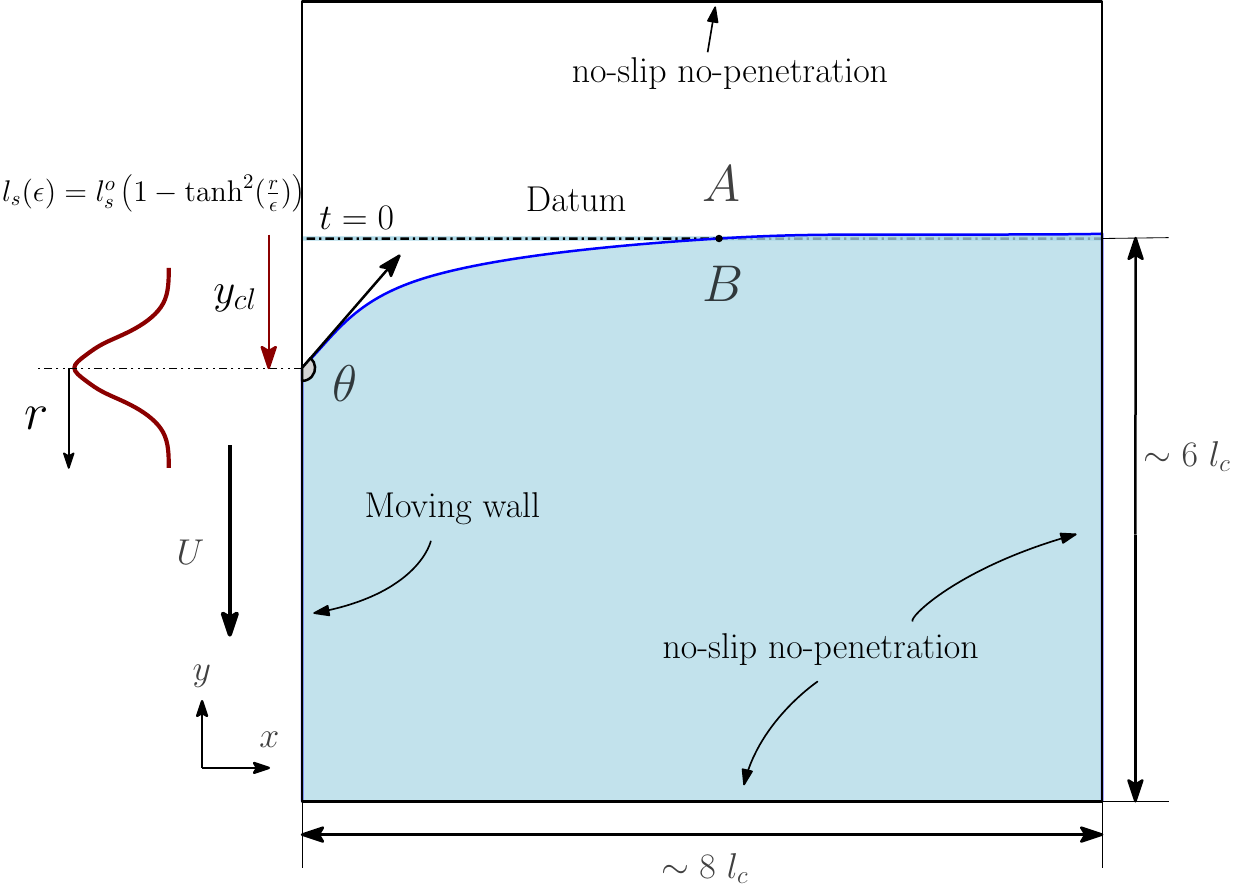}
    \caption{Schematic showing the geometry and boundary conditions for numerical simulations. A variable-slip, $l_s(\epsilon)$, is applied along the moving wall (left boundary) in the vicinity of the contact line which is at a distance $y_{\text{cl}}$ from the initial horizontal level of the fluid, $y=0$.}
    \label{fig:Schematic_sims}   
\end{figure}
\begin{figure}[h]
    \centering
        \includegraphics[trim=0mm 15mm 0mm 5mm, clip, width=0.55\textwidth]{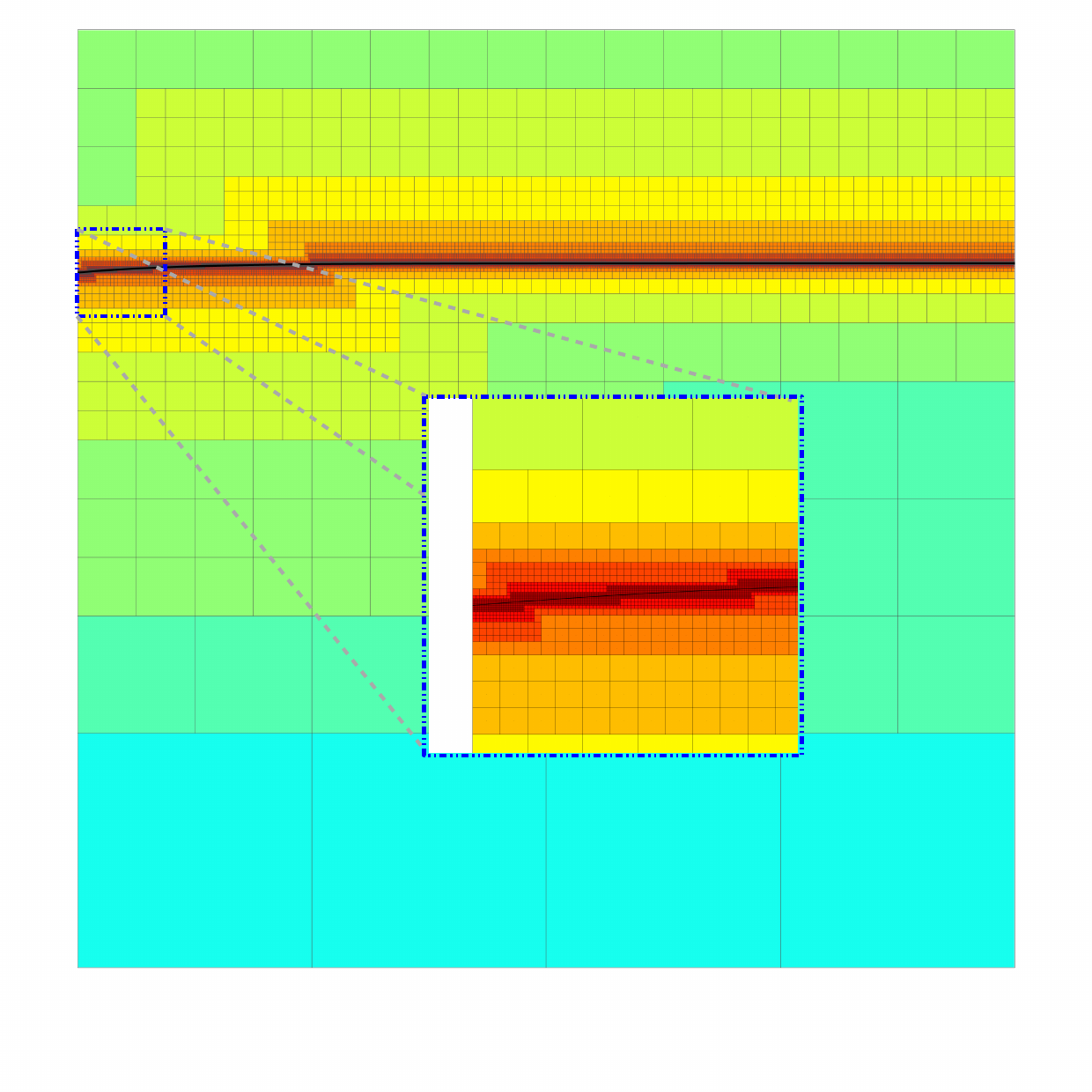}
        \label{1}
    \caption{A typical mesh used in the numerical simulations colored based on the refinement level. A fine mesh is employed along the interface that dynamically adapts to the motion of the interface. In most cases, the smallest mesh size is $10\mu$m obtained at a refinement level of 11. The inset shows a magnified view near the contact line region.}
    \label{Fig:Mesh_2figs}
\end{figure}

The numerical setup closely mimics the experimental domain, but is restricted to two dimensions. A schematic view of the computational setup is shown in figure \ref{fig:Schematic_sims}. The left boundary represents a moving wall where no-slip condition is employed except in a small region in the vicinity of the contact line where a variable slip model is used. More details about the slip model are given later. The side and bottom boundaries are placed at a minimum of $8l_c$ and $6l_c$ respectively while the top boundary was at distance of $2 l_c$. These values were chosen after a convergence study and were found sufficient to eliminate the effect of far-field boundaries. No-slip boundary conditions were imposed on these two boundaries and these conditions had no significant effect on the flow field in the vicinity of the moving contact line. A contact angle, $\theta$, is specified on the left (moving) wall whose values are taken from the experimental data, while an angle of $pi/2$ was imposed on the right wall. It is useful to note that the grid resolution at the contact line in the simulations was comparable to the imaging resolution in the experiments. Thus, the apparent contact angle in the experiments could be directly used as a Dirichlet boundary condition at the moving wall in the simulations. This ensured that the interface shape obtained in numerical simulations closely resembles the experimental findings.
We employ \textit{Basilisk}, a popular open-source solver developed by Popinet and coworkers \cite{popinet2009accurate,popinet2015quadtree,popinet2015collaborators} to solve the Navier-Stokes equations, \eqref{eqn:Governing_eqns}, using the volume-of-fluid method. %
\begin{align}
   \tilde{\bnabla} \cdot \tilde{\bu} = &~ 0 \\
   \rho \left( \dpopa{\tilde{\bu}}{\tilde{t}} + \tilde{\bu} \cdot \tilde{\bnabla}  \tilde{\bu} \right) = &~ - \tilde{\bnabla} \tilde{p} + \bnabla \cdot\left [ \mu \left (\bnabla {\bf u} + \bnabla {\bf u}^T \right )\right ]   + 2 \sigma \kappa \delta_s \mathbf{n} 
   \label{eqn:Governing_eqns}
\end{align}
\textit{Basilisk} employs a second-order accurate Bella-Collela-Glaz (BCG)\cite{bell1989second} advection scheme, which suppresses spurious oscillations in the velocity field. Furthermore, interface normals and curvature are calculated using height functions\cite{afkhami2008height,afkhami2018transition}, which allows for accurate prescription of the contact angle boundary conditions. To ensure accuracy of the interface shape, an adaptive mesh refinement, with a maximum refinement level of 11, is employed in the simulations as shown in figure \ref{Fig:Mesh_2figs}. The smallest grid size in the present simulations is approximately $10\mu$m comparable to the pixel resolution in the experiments.
As described in section \ref{sec:theory_background}, the singularity at the moving contact line with the conventional no-slip boundary condition leads to convergence issues in numerical solutions \cite{afkhami2009mesh}. In a recent study by Fullana \emph{et al.}\cite{fullana2024consistent}, significantly lower grid sizes, in the range of tens of nanometers, were employed at the moving contact line. Such simulations, while being prohibitively expensive, are meant to provide a comprehensive framework to accurately capture the physics of moving contact lines across the entire range of scales present in the problem, i.e. from the microscopic scales where the contact line singularity is alleviated, to the macroscopic scales determining the problem geometry. In their approach, the contact angle is back-calculated based on the velocity field from the neighbouring cells. Such an approach has a universal appeal since it does not require any experimental support for the implementation. In the present study, we adapt the method of Fullana \emph{et al.}\cite{fullana2024consistent}, but suitably modify their approach, allowing us to provide a Dirichlet-type contact angle boundary condition based on experimental inputs. This approach allows us to circumvent the very fine grid sizes to capture microscopic physics near the contact line, and also enables us to compare the simulations and experiments in a coherent manner. 

To establish grid-independence and alleviate the stress singularity, we employ the Navier-slip boundary condition at the contact line with a spatially-decaying localised slip coefficient given by
\begin{equation}
 v_s =  l_s(\epsilon)  \frac{\partial u}{\partial y}, 
 \label{eq:slip_velocity}
\end{equation}
where $v_s$ is the slip velocity at the wall, $\displaystyle l_s(\epsilon) = l_{s,0}\left[1-\tanh^2 \left(\frac{y-y_{\text{cl}}}{\epsilon} \right )\right]$ with $l_{s,0}$ being the value of slip coefficient at the contact line, i.e. at $y=y_{\text{cl}}$, and $\epsilon$ determines the extent of the slip, which is set to 100 $\mu$m in the present study. This value is based on our recent experiments \cite{gupta2024experimental} where it was found that the extent of slip can be substantially large, in the order of 100 $\mu$m in certain conditions, even though the value of the slip coefficient is small. To implement a localised slip model, it is imperative to accurately locate the contact line position on the moving left wall. At each time-step, height functions are used to find the contact line position $y_{\text{cl}}$ from the datum along the left boundary/moving wall. A spatially decaying slip coefficient, modelled as a smeared delta function, is set such that the maximum slip is at the contact line by shifting the delta function to the contact line for every time-step. The localised slip ensures there is maximum slip at the contact line which decays to zero over the distance $\epsilon/2$, ensuring a no-slip boundary condition everywhere else in the vicinity of the contact line, as shown by the red curve in the figure \ref{fig:Schematic_sims}.

Since the simulations in the present study are primarily meant to complement the experiments, physical parameters and the corresponding dimensionless groups such as Reynolds and capillary numbers were chosen based on experimental conditions.

\section{Results}
\label{sec:results}
The main results of this study are presented in three distinct but interlinked parts. In \S\ref{sec:interface_shape}, we compare the interface shapes obtained from experiments with the theoretical predictions described in \S\ref{sec:Interface_Shape_theory}. In \S\ref{sec:flow_fields_ex_MWS}, velocity fields from the PIV experiments are shown as streamfunction contours to enable a direct comparison with theoretical models. Streamfunction contours obtained from numerical simulations are also included for reference and are compared with theory. A direct one-to-one comparison between experiments and simulations is avoided because small but measurable differences in the interface shapes make such a comparison less meaningful. Finally, in \S\ref{sec:interfacial_speed}, the tangential velocity along the interface is extracted from the flow-field data and compared with theoretical predictions. Both experimental and numerical results are presented in this section.{\textcolor{Orange}{[Recheck if needed.]}} A key observation is the gradual slowing down of fluid elements near the contact line. This deceleration is consistent with the presence of a narrow slip region that allows the fluid to turn and align smoothly with the moving plate under finite acceleration.

\subsection{Interface shape comparison}\label{sec:interface_shape}
\begin{figure}[h]
\centering
\subfigure[]{
\label{fig:comparison_3mm_speed_Si_500}
\includegraphics[trim = 0mm 0mm 0mm 0mm, clip, angle=0,width=0.48\textwidth]{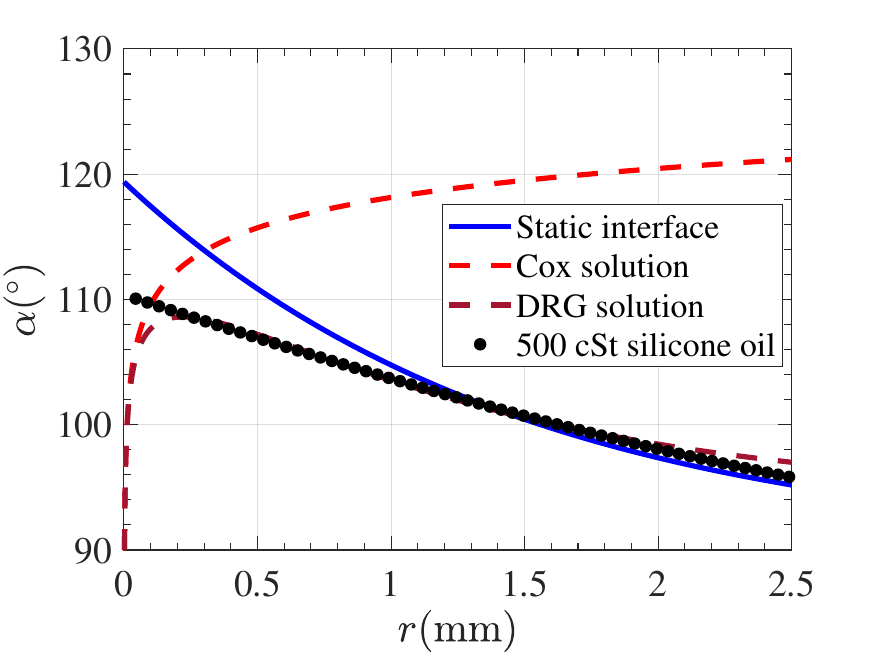}}
\subfigure[]{
\label{fig:comparison_150microns_speed_SW60}
\includegraphics[trim = 0mm 0mm 0mm 0mm, clip, angle=0,width=0.48\textwidth]{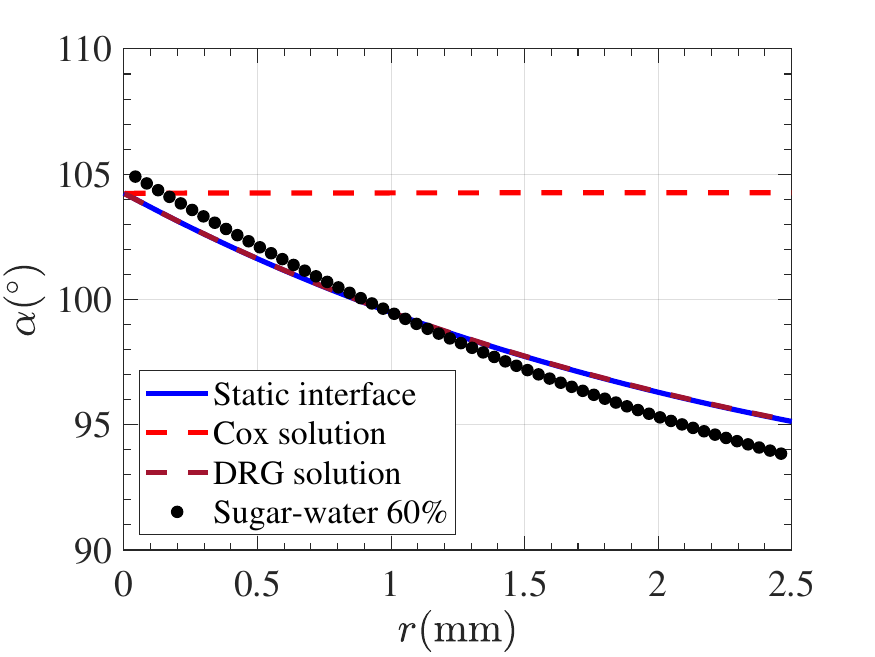}}
\subfigure[]{
\label{fig:comparison_DRG_sugar48_coated}
\includegraphics[trim = 0mm 0mm 0mm 0mm, clip, angle=0,width=0.48\textwidth]{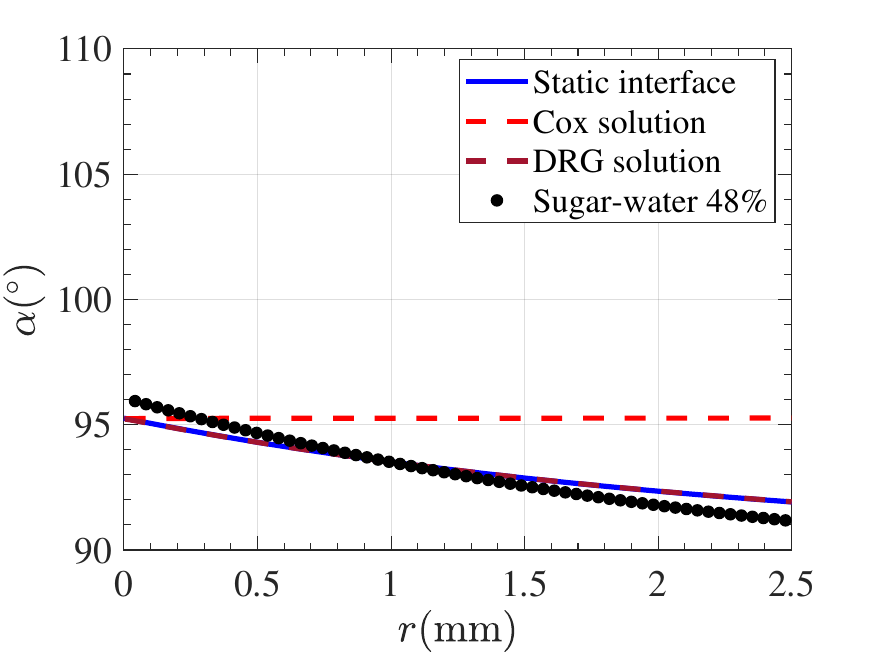}
}
\subfigure[]{
\label{fig:comparison_DRG_water_coated}
\includegraphics[trim = 0mm 0mm 0mm 0mm, clip, angle=0,width=0.48\textwidth]{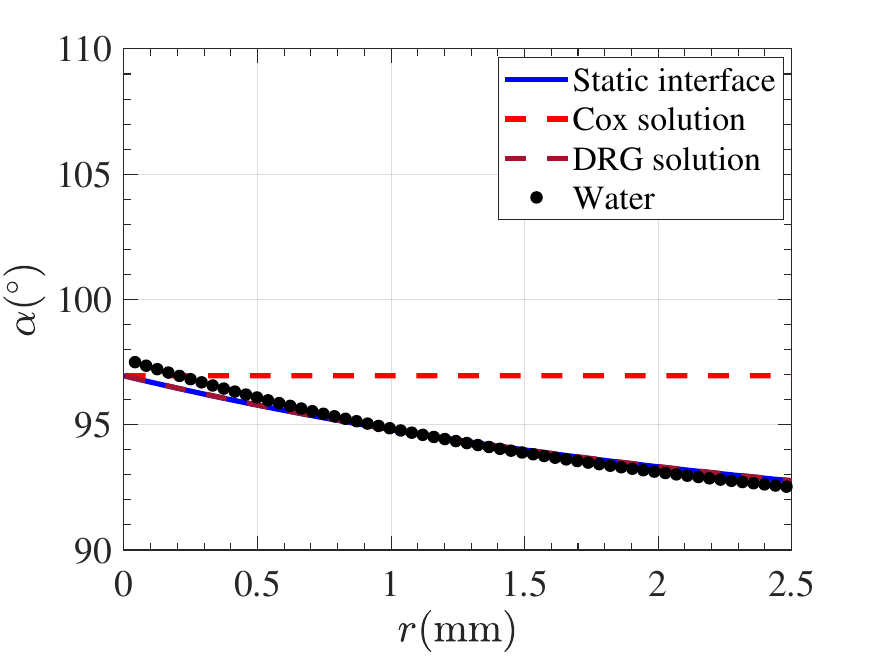}
}
\caption{A comparison of interface shapes among experimental interface shape (dotted black curve), DRG model (maroon dashed curve), Cox model (red dashed curve), and static shape (blue solid curve). (a) air-500 cSt silicone oil at $Ca = 8.37\times 10^{-2}$, $\omega_0 = 119.4$; (b) air-sugar 60\% at $Ca = 1.04 \times 10^{-4}$, $\omega_0 = 104.3$ over a hydrophobic coated solid surface. (c) air-sugar 48\% at $Ca = 8.16\times 10^{-5}$, $\omega_0 = 95.3$ over a hydrophobic coated solid surface. (d) air-water at $Ca = 1.24 \times 10^{-5}$, $\omega_0 = 96.9$ over a hydrophobic coated solid surface.}
\label{fig:interface_shape_DRG}
\end{figure}
\begin{figure}[h]
\centering
\subfigure[]{
\label{fig:comparison_3mm_speed_Si_500_GLM}
\includegraphics[trim = 0mm 0mm 0mm 0mm, clip, angle=0,width=0.48\textwidth]{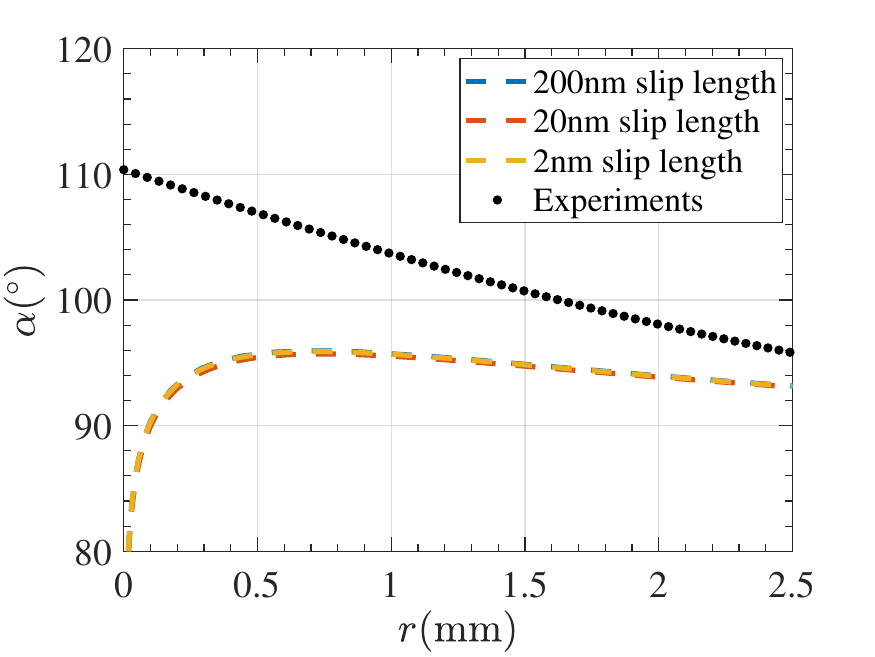}}
\subfigure[]{
\label{fig:comparison_150microns_speed_SW60_coated_GLM}
\includegraphics[trim = 0mm 0mm 0mm 0mm, clip, angle=0,width=0.48\textwidth]{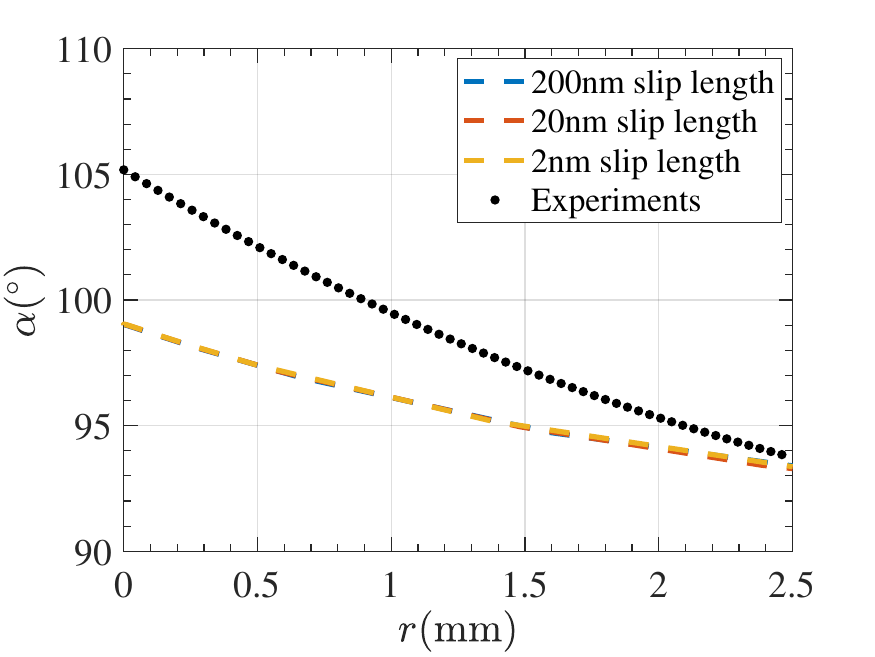}}
\subfigure[]{
\label{fig:comparison_GLM_sugar48_coated}
\includegraphics[trim = 0mm 0mm 0mm 0mm, clip, angle=0,width=0.48\textwidth]{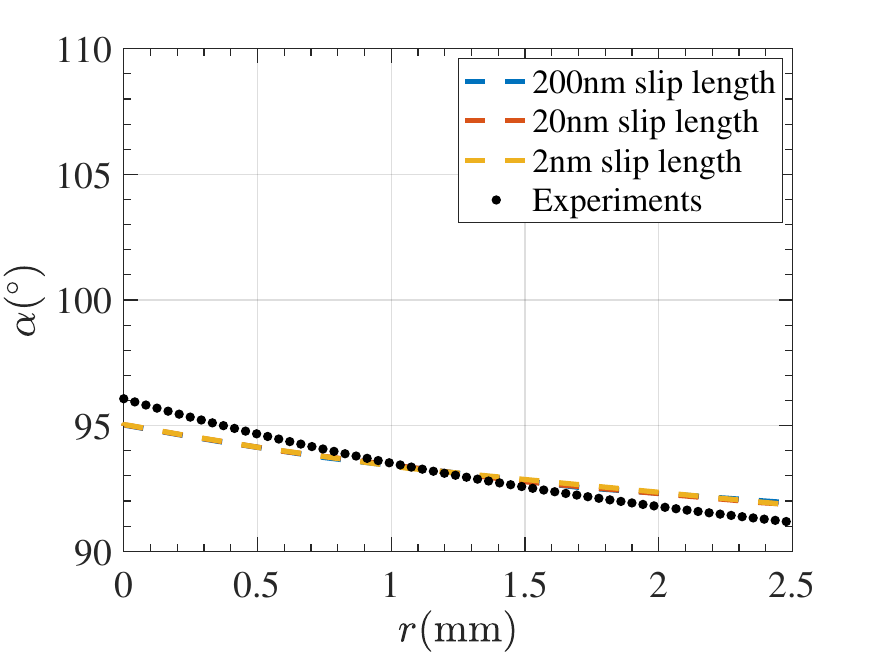}
}
\subfigure[]{
\label{fig:comparison_GLM_water_coated}
\includegraphics[trim = 0mm 0mm 0mm 0mm, clip, angle=0,width=0.48\textwidth]{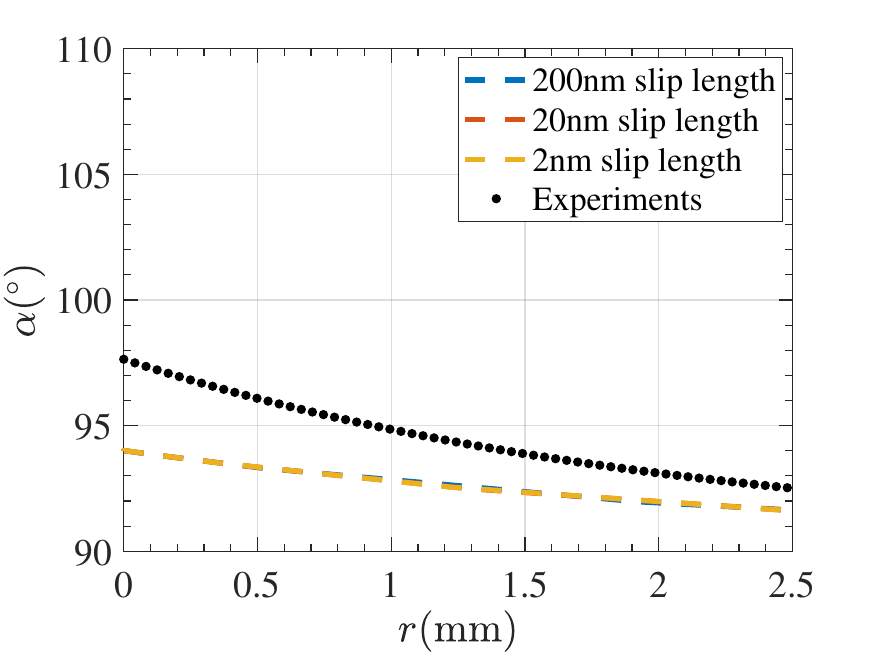}
}
\caption{A comparison of interface shapes between experimental interface shape (dotted black curve) and GLM (dashed curve) with three different slip length 200nm, 20nm and 2nm. (a) air-500 cSt silicone oil at $Ca = 8.37\times 10^{-2}$, $\theta_e = 12$; $c=2.95$; (b) air-sugar 60\% at $Ca = 1.04 \times 10^{-4}$, $\theta_e = 99$; $c=0.85$ over a hydrophobic coated solid surface. (c) air-sugar 48\% at $Ca = 8.16\times 10^{-5}$, $\theta_e = 95$; $c=0.97$ over a hydrophobic coated solid surface. (d) air-water at $Ca = 1.24 \times 10^{-5}$, $\theta_e = 94$; $c=0.99$ over a hydrophobic coated solid surface.}
\label{fig:interface_shape_GLM}
\end{figure}
After the flow has attained a steady state, the interface shape and flow fields are extracted from the PIV data and compared against theoretical predictions. As discussed in section \ref{sec:Interface_Shape_theory}, the DRG model is obtained by matching the `outer' static shape with the inner `Cox' solution while the GLM only requires the values of the slip length and the contact angle as inputs. 

In figure \ref{fig:interface_shape_DRG}, the interface shape from the experiments are compared against the predictions of DRG model for four different values of $Ca$. At high $Ca$, shown in figure \ref{fig:comparison_3mm_speed_Si_500}, since viscous effects are felt farther away from the moving wall, the Cox model shows substantial `viscous bending' near the contact line. The large viscous effects also cause the experimental interface shape to deviate from the static shape, and this is often referred to as `viscous deformation'. The composite DRG solution provides a satisfactory agreement with experimental results except very close to the contact line. The performance of the DRG model improves when the value of $Ca$ is reduced, as is done progressively in figures \ref{fig:comparison_150microns_speed_SW60}, \ref{fig:comparison_DRG_sugar48_coated}, and \ref{fig:comparison_DRG_water_coated}. At very low $Ca$, viscous effects are largely confined to very small regions near the contact line, thus the interface shape is largely governed by the static shape as is clearly evident in figure \ref{fig:interface_shape_DRG}.
In the low $Ca$ limit, the static shape and the interface shape predicted by the DRG model closely follow the experimental interface shape with an error of $\pm 1^{\circ}$. Therefore, the algebraic parameter $\omega_0$ in the DRG model will be identical to the dynamic contact angle $\theta_d$. When the experimental and DRG interface shapes are not in perfect agreement, then $\omega_0$ differs from $\theta_d$. 

The second approach, the GLM, employs a more direct approach to predict the complete interface shape, as expressed in equations \eqref{eq:glm}. This model requires the slip length $l_s$ as an input parameter, in addition to the constant $c$ that depends on the equilibrium contact angle as shown in figure \ref{fig:c_vs_theta}. Since the slip length is unknown, it can be determined by minimizing the deviation between the model's predicted shape and the experimental interface shape. It is important to note that the equilibrium angle, $\theta_e$ need not be equal to Young's angle, and in the present case, $\theta_e$ is considered equivalent to the static advancing contact angle ($\theta_{sa}$) obtained from direct measurements given in table \ref{tab:operating_parameters}.
The comparison of interface shapes from the GLM with experimental results is shown in figure \ref{fig:interface_shape_GLM} for three different slip lengths: 2 nm, 20 nm, and 200 nm. Despite two orders of magnitude variation in slip lengths, the theoretical prediction did not produce any observable deviation in the interface shape. 

Figures \ref{fig:comparison_3mm_speed_Si_500_GLM} and \ref{fig:comparison_150microns_speed_SW60_coated_GLM} indicate that GLM predictions deviate from experimental results, particularly at high $Ca$ when the equilibrium contact angle is imposed as a boundary condition. The GLM model exhibits pronounced bending of the interface, attributed to the transition from the dynamic contact angle to the equilibrium contact angle at the wall, a phenomenon commonly referred to as viscous bending. However, our experiments reveal no such evidence of viscous bending, at least not at the length scales predicted by GLM. In contrast, for low $Ca$ cases, such as water and sugar-water mixtures, the discrepancy between the dynamic and equilibrium angles is relatively small, leading to better agreement between GLM predictions and experimental observations, as shown in figures \ref{fig:comparison_GLM_sugar48_coated} and \ref{fig:comparison_GLM_water_coated}.

Since no viscous bending is visible in the experiments, it is prudent to apply the dynamic contact angle as the relevant boundary conditions in GLM. This requires changing $\theta_e$ to $\theta_d$ in equn. \ref{eq:glmbc}. This leads to a substantial improvement in the agreement between experiments and GLM predictions as discussed in \S \ref{appx:GLM_dynamic} and figure \ref{fig:interface_shape_GLM_thetaD}.

\subsection{Comparisons of flow fields}\label{sec:flow_fields_ex_MWS}
\begin{figure}
\centering
\subfigure[]{
\label{fig:streamfunction_ex_MWS_500cst_Si_2mm}
\includegraphics[trim = 0mm 0mm 0mm 0mm, clip, angle=0,width=0.43\textwidth]{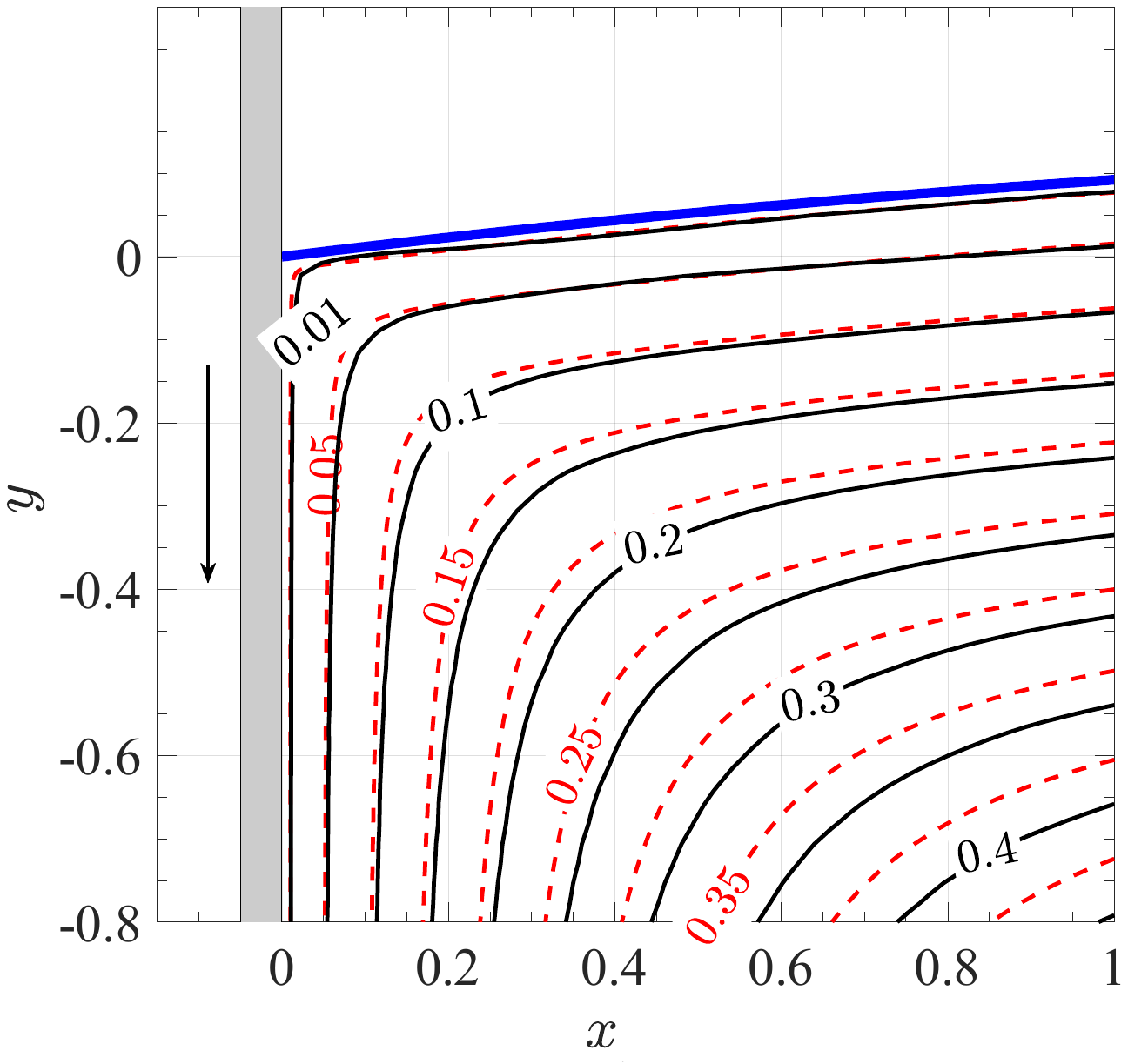}}
\hspace{5mm}
\subfigure[]{
\label{fig:streamfunction_ex_MWS_500cst_Si_3mm}
\includegraphics[trim = 0mm 0mm 0mm 0mm, clip, angle=0,width=0.43\textwidth]{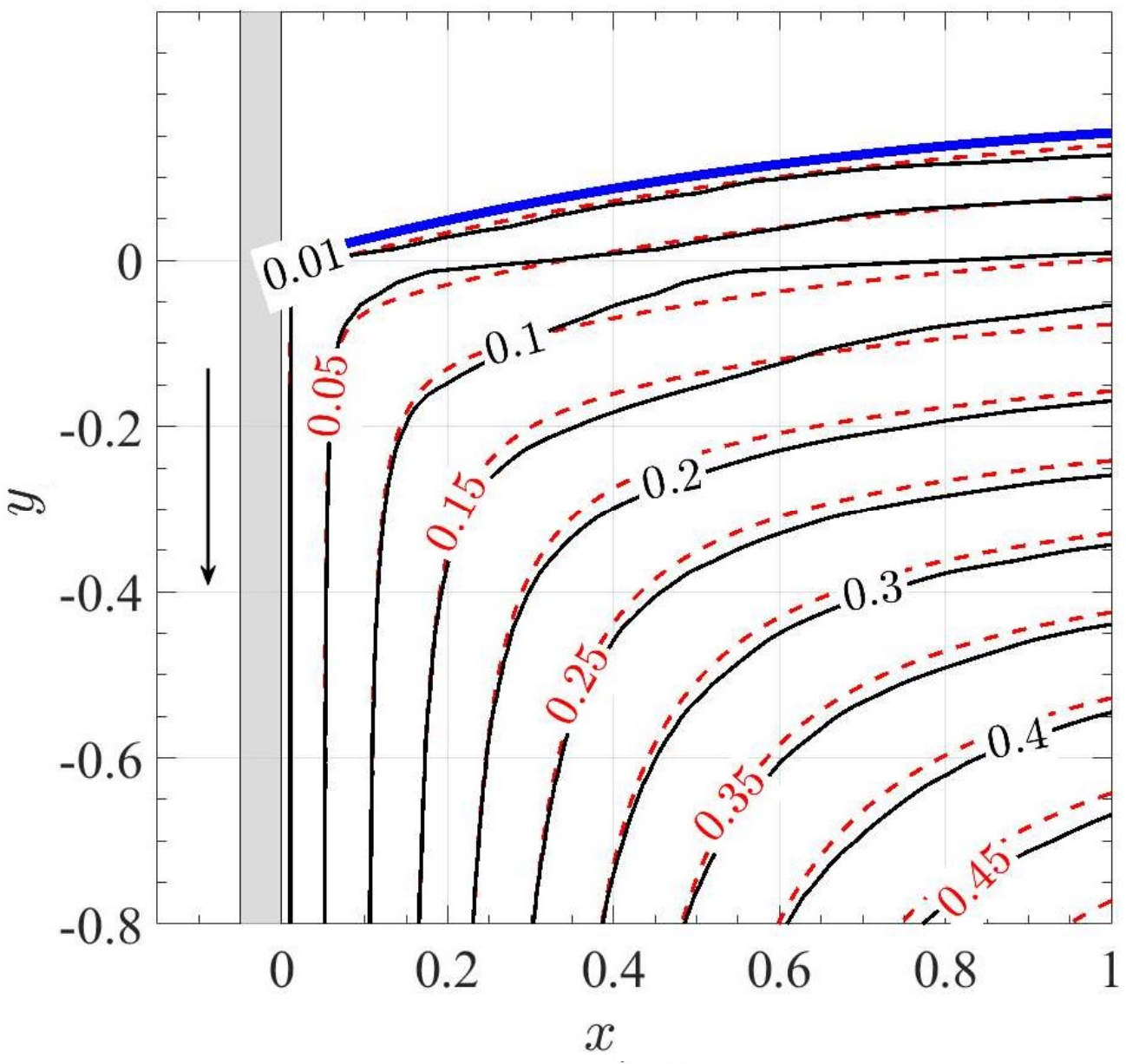}
}
\subfigure[]{
\label{fig:streamfunction_ex_MWS_sugar_coated}
\includegraphics[trim = 0mm 0mm 0mm 0mm, clip, angle=0,width=0.43\textwidth]{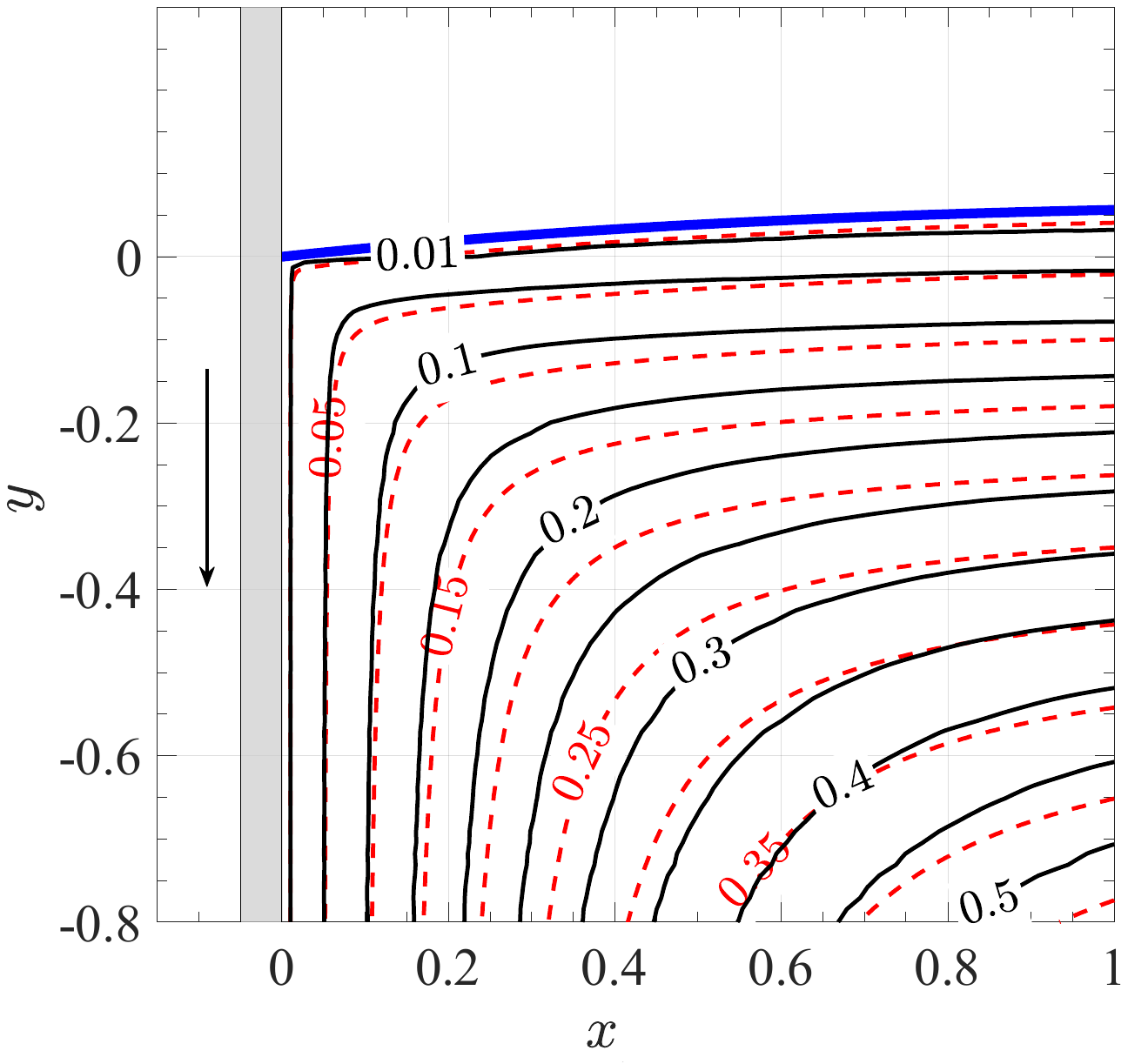}}
\hspace{5mm}
\subfigure[]{
\label{fig:streamfunction_ex_MWS_water_coated}
\includegraphics[trim = 0mm 0mm 0mm 0mm, clip, angle=0,width=0.43\textwidth]{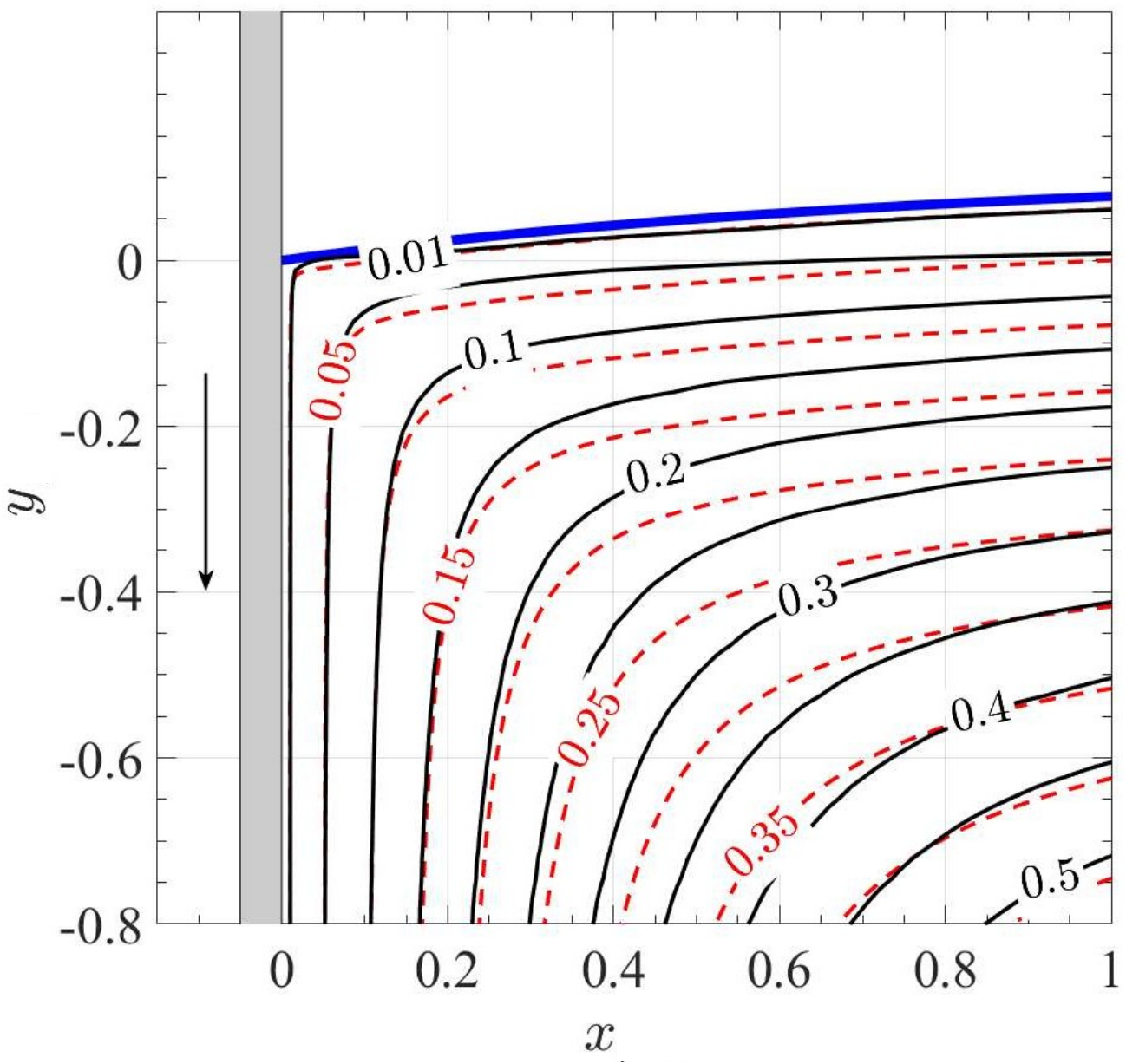}}
\caption{[Experiments] A comparison between streamfunctions of experiments and MWS theory: (a) 500cSt silicone oil $Re = 5.23 \times 10^{-3}$ and $Ca = 5.58 \times 10^{-2}$. 
(b) air-sugar-water mixture 60 \% (w/w) for coated solid surface at $Re= 8.92 \times 10^{-3}$ and $Ca = 1.04 \times 10^{-4}$.
(c)  with air-sugar-water mixture 48 \%  (w/w) for coated solid surface at $Re= 0.125$ and $Ca = 8.16\times 10^{-5}$. (d) air-water for the coated solid surface at $Re= 3.03$ and $Ca = 1.24\times 10^{-5}$ . Dashed curves are theoretical predictions with levels in red colors. The black solid curves are from experimental data with levels in black color. }
\label{fig:expts_obtuse_angle_comparison}
\end{figure} 
The primary objective of the present study is to examine the flow configurations emerging near the moving contact line. This section aims to compare the flow fields from experiments and the corresponding numerical simulations against the MWS theory. Recall that MWS theory is a modification of Huh \& Scriven's theory adapted to a curved interface. The local interface angle, $\beta$ as shown in figure \ref{fig:coordinate_system} is extracted from experiments and numerical simulations. The angle variation $ \beta$ is used as input for MWS theory as explained in \S\ref{sec:theory_background} to obtain streamfunction contours for the corresponding interface shape. This provides the exact form of the streamfunction in a wedge with a curved interface.

\begin{figure}
\centering
\subfigure[]{
\label{fig:MWS_comp_500cSt_SIM}
\includegraphics[trim = 0mm 0mm 0mm 0mm, clip, angle=0,width=0.38\textwidth]{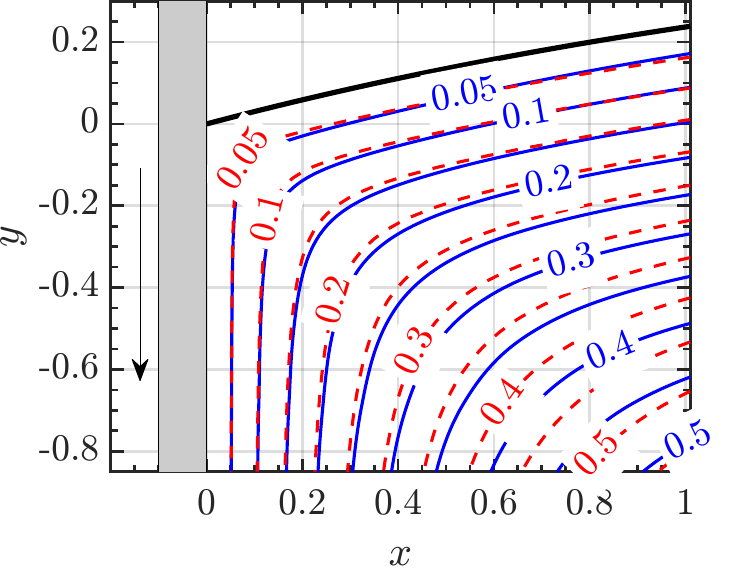}}
\hspace{5mm}
\subfigure[]{
\label{fig:MWS_comp_60cSt_SIM}
\includegraphics[trim = 0mm 0mm 0mm 0mm, clip, angle=0,width=0.38\textwidth]{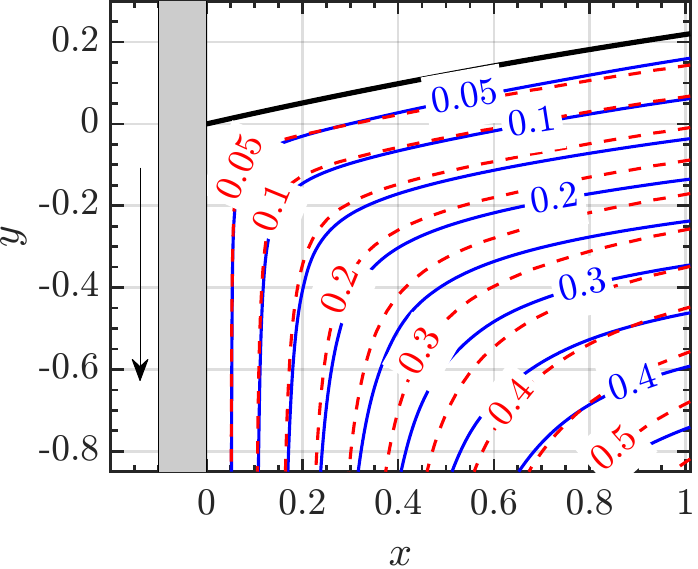}
}
\subfigure[]{
\label{fig:MWS_comp_48cSt_SIM}
\includegraphics[trim = 0mm 0mm 0mm 0mm, clip, angle=0,width=0.38\textwidth]{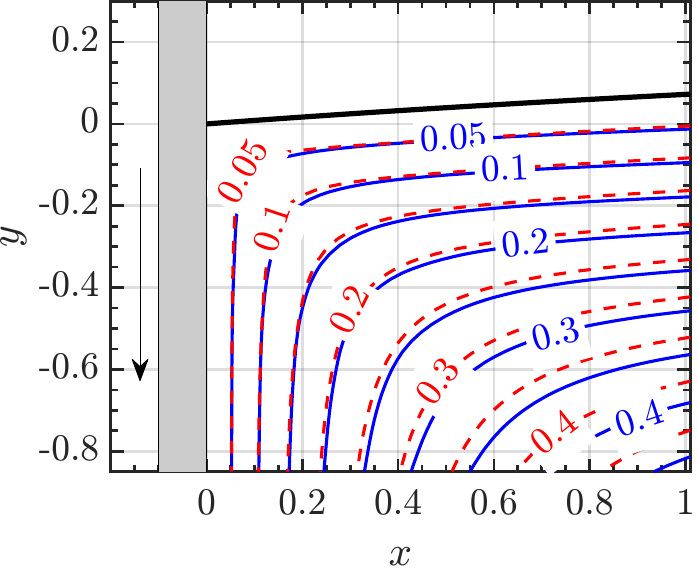}}
\hspace{5mm}
\subfigure[]{
\label{fig:MWS_comp_AW_SIM}
\includegraphics[trim = 0mm 0mm 0mm 0mm, clip, angle=0,width=0.38\textwidth]{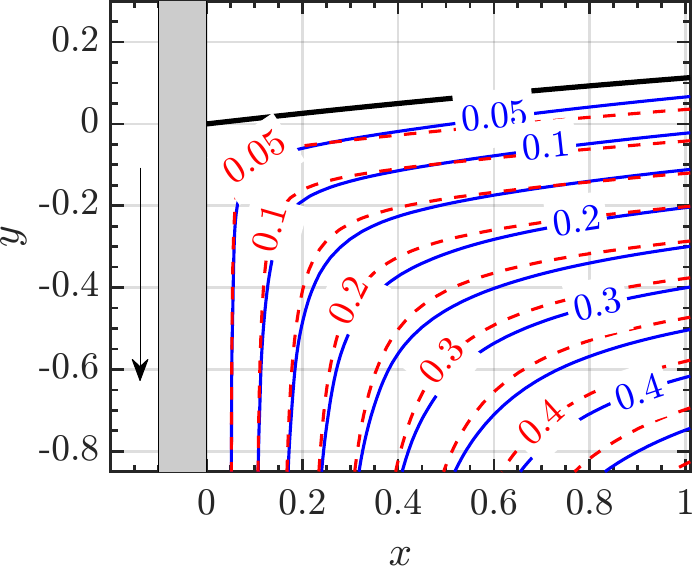}}
\caption{[Simulations] A comparison between streamfunctions from simulations and MWS theory: 
(a) 500~cSt silicone oil at $Re = 5.23 \times 10^{-3}$ and $Ca = 5.58 \times 10^{-2}$. 
(b) Air--sugar--water mixture (60\% w/w) for a coated solid surface at $Re = 8.92 \times 10^{-3}$ and $Ca = 1.04 \times 10^{-4}$. 
(c) Air--sugar--water mixture (48\% w/w) for a coated solid surface at $Re = 0.125$ and $Ca = 8.16 \times 10^{-5}$. 
(d) Air--water system for a coated solid surface at $Re = 3.03$ and $Ca = 1.24 \times 10^{-5}$. Red dashed curves represent theoretical predictions with levels in red labels, while blue solid curves are obtained from simulation data with levels indicated in blue.}
\label{fig:sims_obtuse_angle_comparison}
\end{figure} 

The comparisons are performed by overlaying streamfunction contours obtained from experiments and the MWS theory. In a similar fashion, streamfunction contours from numerical simulations are plotted against MWS theory. In the case of experiments, particle images from PIV are post-processed to obtain velocity field on a uniform grid. The corresponding streamfunction field is then obtained by calculating the mass flux at each consecutive pair of nodes in the flow field data matrix. This procedure is described in more detail in appendix \S\ref{appx:streamfunction_calculation}. In the case of numerical simulations, vorticity field, $\Omega$, is directly obtained from the velocity which in turn is used to determine streamfunction as a solution of a Poisson equation, i.e. $\bnabla^2\psi = -\Omega$. The value of the streamfunction for MWS theory was obtained directly from the streamfunction expression mentioned in equn. \eqref{eq:Modulate_wedge_1}. 
The slight differences in the interface shapes between experiments and simulations produces noticeable variations in the shapes of the streamfunction contours. As a result, two distinct comparisons are made for experiments and simulations with MWS theory. 
Such a direct comparison of streamfunction contours provides the most rigorous test of theoretical predictions with experimental data, and also demonstrates the efficacy of numerical methods to capture flow fields accurately in the vicinity of the moving contact line. It is important to note that no fitting parameters are involved in this comparison, except for the input of interface shape in the theory.

Figure \ref{fig:expts_obtuse_angle_comparison} shows streamfunction contours from experiments with four different fluids, viz., 500cSt silicone oil, 60\% (w/w) sugar-water mixture, 48\% (w/w) sugar-water mixture, and clean water. The accuracy of the theoretical model can be assessed from the agreement of contour shapes and contour levels with experiments. Huh \& Scriven's wedge flow model (presented in the form of MWS) is in fair agreement with experimental observations.
In all cases reported in this study, material points at the interface migrate toward the contact line and subsequently align with the moving solid. Similar results were also obtained for all the other liquid-gas combinations given in Table \ref{tab:operating_parameters}.

Figures \ref{fig:streamfunction_ex_MWS_500cst_Si_2mm} and \ref{fig:streamfunction_ex_MWS_500cst_Si_3mm} illustrate remarkable agreements between MWS theory and experiments for 500cSt silicone oil and 60 \% sugar-water mixture, respectively. Here, experimental data are depicted by black solid curves, and MWS theory is depicted by red dashed curves. The comparison is good near the contact line, but deviations become evident as the distance from the contact line increases. Figures \ref{fig:streamfunction_ex_MWS_sugar_coated} and \ref{fig:streamfunction_ex_MWS_water_coated} present streamfunction comparisons for 48\% sugar-water mixture and water. Here, the contour levels associated with experimental streamfunctions deviate from the contour levels associated with theoretical streamfunctions. The discrepancies are minimal near the moving solid. However, they become more pronounced as one moves away from the solid and the contact line. Additionally, the experimental streamfunctions appear compressed against the solid surface compared to the streamfunctions derived from the MWS theory. This discrepancy is observed in 48\% sugar-water mixture and water systems, both of which are low-viscosity fluids. These deviations can be attributed to the influence of inertial forces. 

In the same spirit as the experiments, figure \ref{fig:sims_obtuse_angle_comparison} depicts streamfunction contours obtained from numerical simulations. The interface shape extracted from the numerical simulations are used to determine theoretical streamfunction contours using MWS theory. Clearly, the streamfunction contours in figure \ref{fig:sims_obtuse_angle_comparison} demonstrates the ability of numerical simulations to capture flow fields both qualitatively and quantitatively. This was possible due to two main ingredients incorporated into numerical simulations as described in \S\ref{sec:numerics}: (i) the incorporation of variable slip at the moving contact line, (ii) ability to prescribe a dynamic contact angle in the simulations to allow nearly identical interface shapes and flow fields. But such an approach relies on prior knowledge of experimental results, thus such simulations can only complement existing experimental data. Again, figures \ref{fig:MWS_comp_500cSt_SIM}-\ref{fig:MWS_comp_AW_SIM} show excellent agreement between numerical simulations and MWS theory. The nearly identical shape of the flow fields in figures \ref{fig:expts_obtuse_angle_comparison} and \ref{fig:sims_obtuse_angle_comparison} shows that numerical simulations can indeed be used as a reliable tool to probe the dynamics in the vicinity of the moving contact line. Furthermore, close agreement in streamfunction values between experiments and simulations reveals that numerical simulations can be used to complement experiments, particularly in those cases where PIV measurements are not feasible. 

\subsection{Interfacial speed comparison} \label{sec:interfacial_speed}
\begin{figure}[h]
\centering
\subfigure[]{
\label{fig:interfacial_speed_with_inset}
\includegraphics[trim = 0mm 0mm 0mm 0mm, clip, angle=0,width=0.7\textwidth]{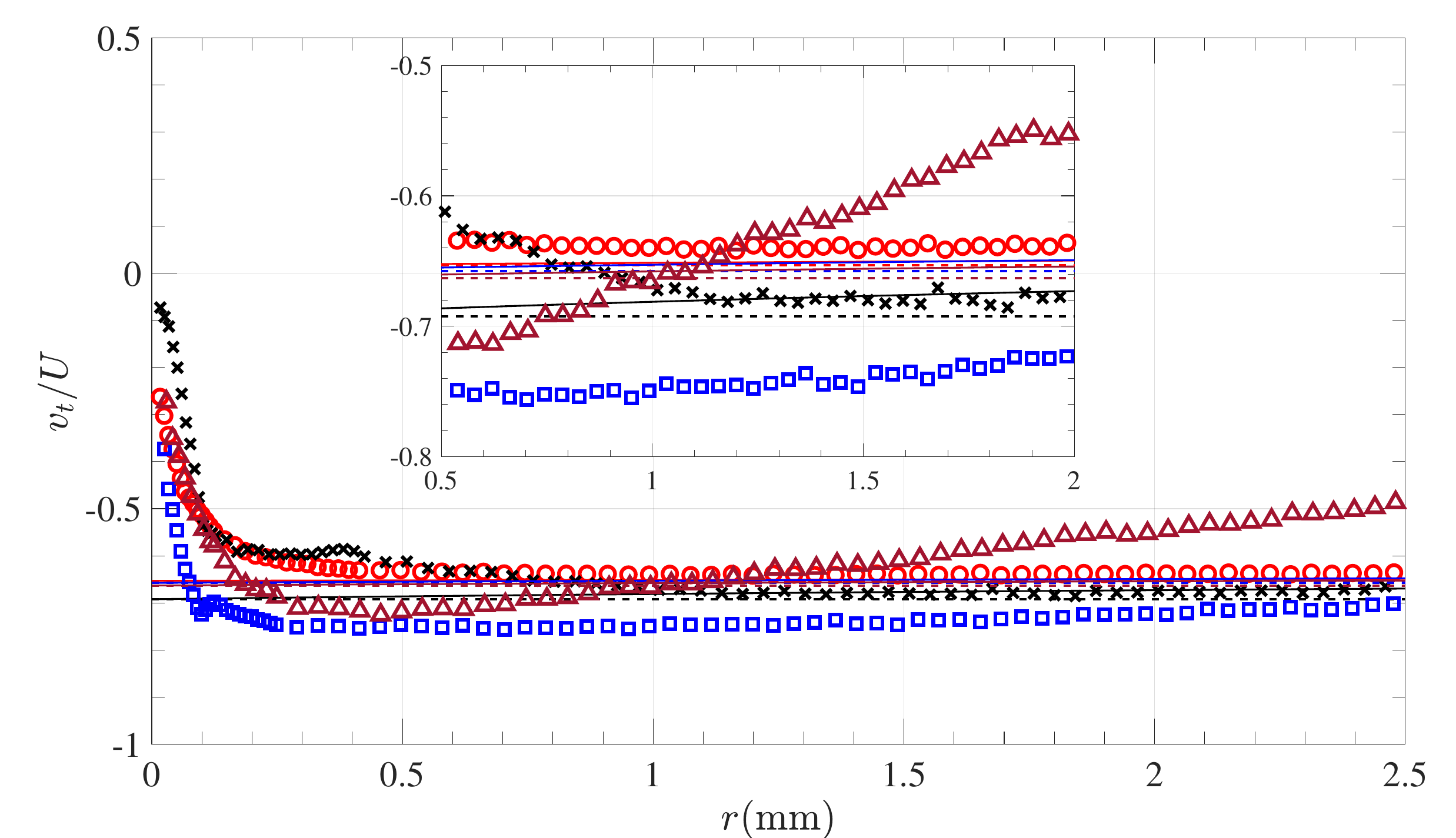}}
\subfigure[]{
\label{fig:interfacial_speed_closeup}
\includegraphics[trim = 0mm 0mm 0mm 0mm, clip, angle=0,width=0.7\textwidth]{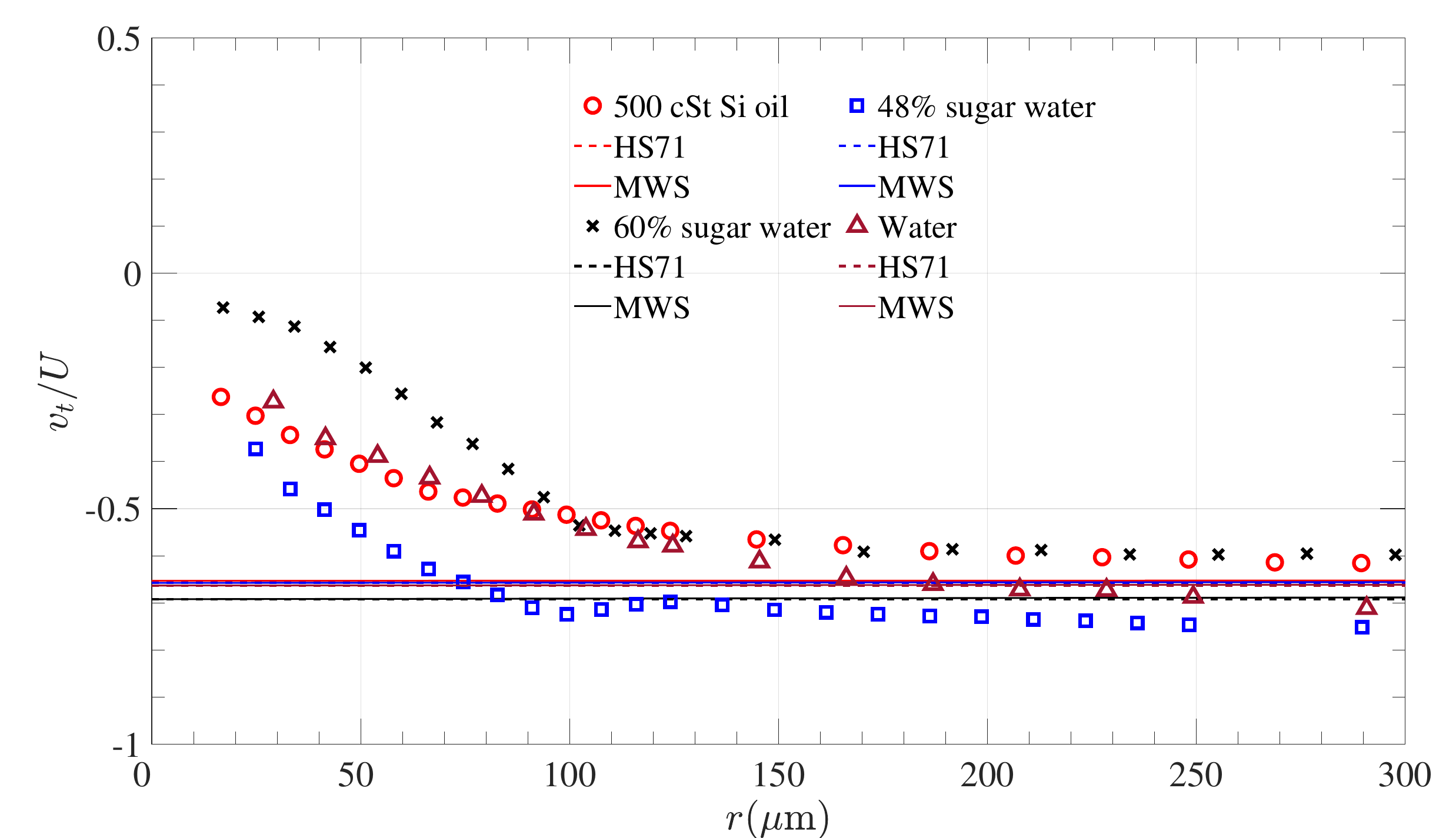}}
\caption{ (a) Non-dimensional tangential speeds ($v_t/U$) for different fluid-fluid systems, where $U$ is the plate speed. The speeds are measured along the interface, varying with respect to $r$. The negative values of the speeds indicate the motion of the interface towards the contact line, implying a rolling-type motion in the fluid domain. The inference of flow patterns from the direction of the speed is consistent with the streamfunction contours shown in figure \ref{fig:expts_obtuse_angle_comparison}. 
Details of experimental data points are as follows:~~ For 500 cSt silicone oils: $\color{red}\boldsymbol{\medcirc}$: $Re = 5.23 \times 10^{-3}$,$Ca = 5.58 \times 10^{-2}$,  $\phi = 96.9^{\circ}$;~~~ For 60\% sugar-water mix: $\color{black}\boldsymbol{\bigtimes}$: $Re= 8.92 \times 10^{-3}$, $Ca = 1.04 \times 10^{-4}$, $\phi = 105^{\circ}$;~~~ For 48\% sugar-water mix: $\color{blue}\boldsymbol{\square}$:$Re = 1.25 \times 10^{-1}$,$Ca = 8.16 \times 10^{-5}$, $\phi = 96.1^{\circ}$;~~ For water: 
$\color{maroon}\boldsymbol{\bigtriangleup}$: $Re = 3.03$,$Ca = 1.24 \times 10^{-5}$, $\phi = 97.6^{\circ}$;.
Interfacial speeds from experiments are denoted with different markers. The predictions from HS71 and MWS theory are shown with dashed lines and solid curves, respectively. The inset shows a close-up view of the interfacial speed away in the far field. (b) A magnified view of interfacial speed near the contact line is shown to emphasize the rapid reduction in the speeds while approaching the contact line.}
\label{fig:interfacial_speed_comparison}
\end{figure} 

The motion of fluid particles along the interface provides insight into the flow behavior in the bulk regions away from the interface. For advancing contact lines, motion of the liquid toward the contact line indicates a rolling motion within the liquid and a split-streamline motion in the gas phase. This approach was first used by Dussan and Davis~\cite{dussan1974motion}, who showed that liquid drops exhibit rolling motion near the advancing contact line. In the present study, interface speeds are obtained from both experiments and numerical simulations and compared with the HS71 and MWS theoretical predictions. Recall from \S\ref{sec:theory_background} that the HS71 theory predicts a constant interfacial speed for a given wedge angle (see Eq.~\ref{eq:Scriven2}), whereas the MWS theory introduces a small variation away from the contact line due to interface curvature. The constant interfacial velocity at the contact line predicted by these models implies that material points at the interface strike the moving wall with finite speed and must instantaneously align with it — yet another manifestation of the classical contact-line singularity.

Here, we provide direct evidence of a variable interfacial speed in both experiments and simulations. In the experiments, the interfacial speed is obtained by projecting the velocity field data along the interface. In numerical simulations, the interfacial position is first identified using the VOF parameter; velocity fields in the corresponding interfacial cells are then extracted, and the interfacial speed is determined by projecting this velocity onto the interface. For consistency, the variation of interfacial speed is shown with respect to the radial position along the interface, denoted by $r$.

Figure~\ref{fig:interfacial_speed_comparison} presents the measured interfacial speeds for the four liquid–air systems discussed in Fig.~\ref{fig:expts_obtuse_angle_comparison}. The normalized interfacial speed $v_t/U$ is plotted as a function of the radial position $r$, with the contact line located at $r = 0$. In all cases, the tangential speed is negative, indicating that fluid elements move toward the contact line. This observation is consistent with a global rolling-type motion in the flow domain, as described by HS71 theory. 

According to HS71 theory, the interfacial speed should remain constant (dotted lines), whereas the MWS theory predicts a gradual decrease in speed along the interface (solid curves). Although the MWS correction accounts for curvature effects, the deviation from HS71 predictions remains small under the present experimental conditions. In all four systems, the experimental results show that away from the contact line ($r > 150~\mu$m), the interface speeds are of the same order of magnitude as those predicted by HS71 theory but deviate significantly closer to the contact line. For 500~cSt silicone oil and the 60\% sugar–water mixture, the agreement with theory is excellent, while for the 48\% sugar–water mixture and water, slightly higher experimental values are observed. These minor discrepancies can be attributed to the hydrophobic coating on the solid plate, which may lead to a microscopic contact angle larger than the macroscopic value measured in the experiments, thereby altering the interfacial velocity relative to theoretical predictions. In the water system, the interfacial speed decreases in magnitude away from the contact line, likely due to the influence of inertia, which becomes significant because of water’s low viscosity. Further investigation is required to fully assess the role of inertia in moving contact-line dynamics.

Figure~\ref{fig:interfacial_speed_closeup} provides a magnified view of Fig.~\ref{fig:interfacial_speed_with_inset}, focusing on the region within $300~\mu$m of the contact line. Within the last $100$–$150~\mu$m, a sharp decline in interfacial speed is evident, contradicting the constant-velocity prediction of HS71. The HS71 theory assumes that the interfacial speed maintains its value up to the contact line, at which point it abruptly changes direction to align with the moving wall, leading to a mathematical singularity. In contrast, our experimental results reveal a rapid but finite decrease in speed before alignment with the plate, suggesting a continuous and physically realizable transition. This pronounced slowing of interfacial motion near the contact line may therefore offer a plausible resolution to the classical contact-line singularity problem.

\begin{figure}[h]
    \centering
    \subfigure[]{
        \includegraphics[trim=0mm 0mm 0mm 0mm, clip, width=0.50\textwidth]{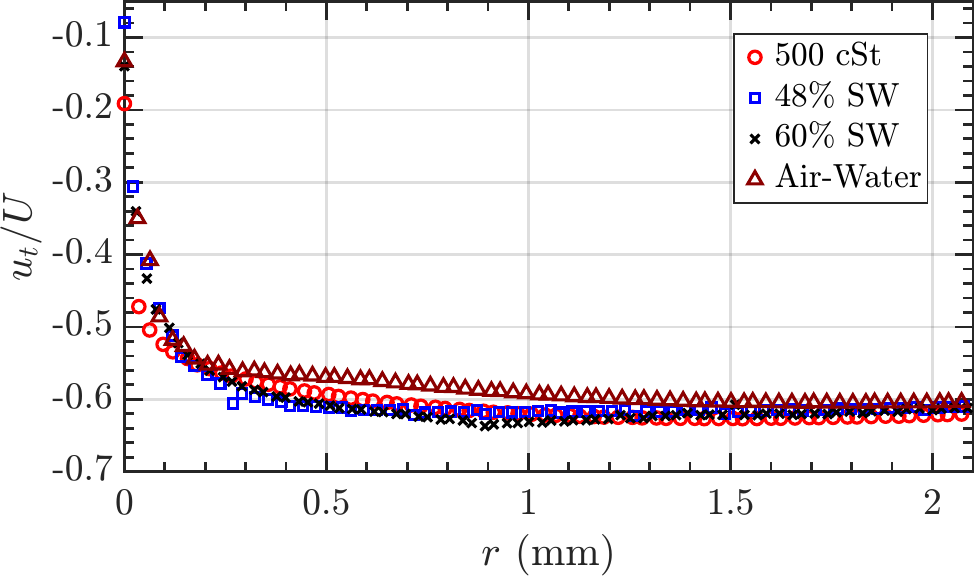}
        \label{fig:int_speed_sims}
    }
    \hspace{0mm}
    \subfigure[]{
        \includegraphics[trim=0mm 0mm 0mm 0mm, clip, width=0.45\textwidth]{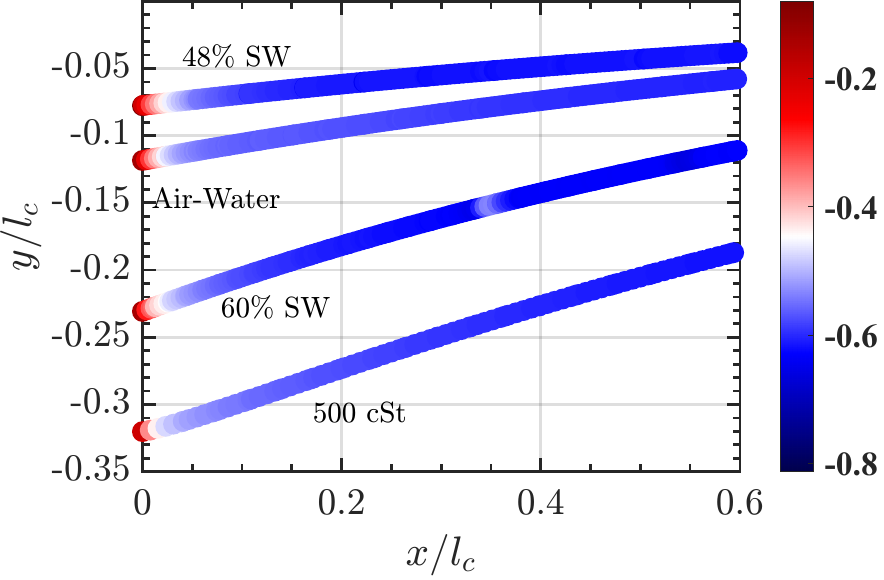}
        \label{fig:int_speed_location_speed}
    }
    \caption{(a) Variation of dimensionless interfacial speed from simulations plotted along the interface. (b) Interface shapes from simulations colored with the interface velocity illustrating the slowing of material points on the interface as they approach the contact line. The four curves correspond to the fluid parameters and dimensionless numbers given in figure \ref{fig:interfacial_speed_comparison}.}     
\end{figure}

To complement the experimental findings, interfacial speeds obtained from numerical simulations are shown in Fig.~\ref{fig:int_speed_sims}. As in the experiments, numerical simulations also show a rapid slowing within approximately $r \leq 300~\mu$m of the contact line. 
The numerical results show consistent behaviour across all systems — a gradual reduction in speed followed by a sharp decrease close to the contact line. The precise location where deceleration begins can be fine-tuned by adjusting the slip-length parameter $\epsilon$ in Eq.~\eqref{eq:slip_velocity}. In Fig.~\ref{fig:int_speed_location_speed}, the interfacial shape is shown colored by the local velocity, illustrating the pronounced slowing of material points as they approach the contact line. The consistency between the location and extent of the slowing region in experiments and simulations confirms that this feature is a robust characteristic of flow near a moving contact line.

\section{Summary and Discussions}
\label{sec:conclusion}

The present study reports simultaneous measurements of flow fields, interface shapes, and interfacial speeds in the vicinity of a moving contact line, with a view to comparing these results against established theoretical models and Volume-of-Fluid (VoF) based numerical simulations. Theoretical frameworks considered here are formulated for the viscous regime, specifically under conditions where $Re \ll 1$ and $Ca \ll 1$. In the present experiments, while the capillary number remained small ($10^{-6} \leq Ca \leq 10^{-2}$), the Reynolds number varied from $10^{-3}$ to $10^{1}$, thereby extending the applicability of the results to low and moderate inertia conditions.

The moving contact line dynamics were investigated by systematically varying the viscosity ratio and dynamic contact angle using different fluid pairs at controlled plate velocities. All experiments were conducted in an advancing plate configuration, which offers significant advantages in minimizing optical distortions and ensuring high-fidelity PIV measurements near the contact line. In contrast to droplet motion on inclined planes or capillary-driven flows in tubes, the present geometry avoids confinement effects, allowing an unambiguous capture of the local flow structure. The working fluids included 500~cSt silicone oil, sugar-water mixtures of varying concentrations, and water, thereby spanning a broad range of viscosities and density ratios. This wide parameter space facilitated a comprehensive examination of viscous, capillary, and inertial contributions to the flow.

The PIV-based velocity fields revealed a well-defined stagnation-type topology near the contact line, consistent with theoretical expectations. The interface curvature and local flow gradients exhibited systematic dependence on both viscosity and plate velocity. As the capillary number increased, the dynamic contact angle increased correspondingly, consistent with the trends predicted by classical hydrodynamic theories. For low-$Ca$ conditions, the flow was predominantly viscous, with interfacial motion governed by local curvature and slip; however, at higher $Re$, inertial effects began to alter the streamline topology and interfacial velocity distribution.

Theoretical models such as the DRG and GLM were employed to predict interface shapes across various length scales. For $Ca \ll 1$, both models provide valuable insights, though with differing fidelity near the contact region. At moderate $Ca$, the DRG model accurately predicts the interface shape away from the contact line, while discrepancies arise close to the contact region, particularly in the 500~cSt silicone oil system where strong viscous effects prevail. At very low $Ca$, the DRG model remains valid across nearly all observable scales in air–sugar–water and air–water systems, as the viscous region near the contact line is confined to micrometric scales, below the resolution of the present optical setup. Consequently, for low $Ca$, the static interface shape parameterized by angle $\omega_0$ is sufficient to reproduce experimental interface shapes. The GLM, in contrast, shows larger deviations at both low and moderate $Ca$, primarily due to its use of the equilibrium contact angle as the boundary condition. For high-viscosity fluids such as 500~cSt silicone oil, where $\theta_{sa} \approx 12^{\circ}$, this assumption leads to significant discrepancies in obtuse dynamic-angle scenarios. The GLM performs moderately well for low-viscosity fluids such as water and dilute sugar-water mixtures with $\theta_{sa}\approx 90^{\circ}$, but even in these cases, the DRG model offers a closer fit to experimental data. Given these limitations, experimental interface shapes were adopted for subsequent streamfunction calculations within the MWS framework.

Comparisons of flow fields were performed using streamfunctions computed from mass conservation between consecutive PIV data points. Streamfunction contours from experiments, VoF simulations, and MWS theory exhibited close correspondence near the contact line, confirming the validity of viscous-flow assumptions in this region. Deviations at larger distances from the contact line are consistent with the breakdown of the local approximations inherent to HS71 theory. The agreement between all three datasets reinforces the physical consistency of the measured and simulated fields.

The interfacial speeds obtained from experiments were projected along the interface and compared against predictions from HS71 and MWS theories. For highly viscous fluids such as 500~cSt silicone oil and 60\% sugar-water mixtures, experimental interfacial speeds matched theoretical expectations away from the contact line, while a rapid decay in speed was observed close to the contact region. This deceleration offers direct experimental confirmation of the classical moving-contact-line singularity problem: whereas HS71 theory predicts a finite contact-line speed and an associated infinite stress, the experiments demonstrate a smooth vanishing of speed as the fluid aligns with the solid. In low-viscosity fluids (48\% sugar-water and air–water systems), the experimental interfacial speeds exceed HS71 predictions away from the contact line, likely due to surface wettability effects. The hydrophobic coating applied to the substrate to maintain obtuse angles may cause the microscopic contact angle to exceed the macroscopic value, thereby enhancing the local interfacial speed. For the air–water system, the interfacial speed exhibits an initial rise followed by decay, indicative of inertial dominance at higher Reynolds numbers.

In addition to the experiments, VoF-based numerical simulations were carried out to complement and interpret the experimental observations, employing identical geometric configurations and material properties. A variable-slip model was incorporated into the simulations to alleviate the classical contact-line singularity and to ensure numerical convergence. The simulations successfully reproduced the global interfacial topology, velocity fields, and contact-line motion observed experimentally. Notably, they captured the gradual deceleration of interfacial velocity in the immediate vicinity of the contact line, in close agreement with the experimental measurements. Consistent with the experiments, the simulations exhibited excellent agreement with viscous theory, particularly at low $Re$. Minor discrepancies in the interfacial shape between experiments and simulations can be attributed to experimental factors such as dynamic contact-angle hysteresis and the limited spatial resolution near the wall. Additionally, the slip parameter, $\epsilon$, could be adjusted to improve quantitative agreement, although such fine-tuning was not the objective of the present work. The simulations were intended primarily to provide independent support for the experimental findings. These results confirm that VoF-based simulations, equipped with variable-slip boundary models, can serve as a robust and predictive tool for analyzing flow dynamics near moving contact lines. The present numerical framework can readily be extended to more complex configurations, including receding contact lines, partial wetting, and wetting–dewetting transitions.

In summary, the investigation demonstrates that the viscous theories of HS71 and MWS provide qualitatively accurate descriptions of the flow, even at moderate $Re$, while the combined experimental and numerical analysis confirms the onset of inertial and wetting-related effects beyond the strictly viscous regime. The close correspondence among theory, experiments, and VoF simulations establishes a unified framework for understanding moving contact lines, and offers a reliable foundation for extending such analyses to three-dimensional, unsteady, and hysteretic wetting phenomena in future studies.

\vspace{5mm}
\begin{acknowledgments}
We wish to acknowledge the support of the Anusandhan National Research Foundation (formerly Science and Engineering Research Board), Department of Science and Technology, Government of India, for funding this research through grant no. CRG/2021/007096.  VSAS thanks the Prime Minister's Research Fellowship for financial support.
\end{acknowledgments}

\vspace{5mm}
\noindent \textbf{DATA AVAILABILITY}\\
The data that support the findings of this study are available from the corresponding author upon reasonable request.

\bigskip

\appendix
\section*{APPENDIX}
\setcounter{section}{0}
\renewcommand{\thesection}{\Alph{section}}
\section{The expression for static interface shape}
\label{appx:static_shape}
A static interface shape for flat plate geometry can be represented in a parametric form as follows:
\begin{equation}
    \theta_s(r) = \frac{\pi}{2} - \tan^{-1}\left(\frac{dh_s}{dx}\right),
  \label{eq:f0}
\end{equation}
%
In the given expression, $\theta_s$ represents a local slope from the vertical direction (see figure \ref{fig:omega_o}), with the subscript $s$ indicating a static solution. $h_s$ and $x$ denote the non-dimensional horizontal and vertical location of an arbitrary point on the static interface, respectively.

The full nonlinear Young-Laplace equation can be written expressing the $h_s$ in terms of $x$ and $x_0$ as
%
\begin{equation}
  h_s(x) = l_c \left[\cosh^{-1}\left(\frac{2l_c}{x_0-x}\right) - \cosh^{-1}\left(\frac{2l_c}{x_0}\right)\right] - \left(4 l_c^2 - (x_0-x)^2\right)^{1/2} + \left(4l_c^2-x_0^2\right)^{1/2},
  \label{eq:static_shape_parametric}
\end{equation}
where
\begin{equation}
  x_0 = \sqrt{2}l_c \left(1-\sin\omega_0\right)^{1/2}. 
\end{equation}
%
Here, $l_c$ is the characteristic length (capillary length in the present study), and $x$ represents the vertical location of the interface, which starts from 0 at the contact line to $x_0$ in the far field. $\omega_0$ denotes an apparent angle measured at $x=0$ and determined through an iterative error minimization process.


\section{The expressions of streamfunction and interfacial speed for fixed wedge geometry} \label{appx:fixed_wedge}
\noindent

Coefficients of streamfunction equation for fluid phase B are as follows:
\begin{align*}
a_1 =& \frac{1}{\sin\beta}(-a_3\beta\sin\beta - a_4\beta\cos\beta) \\[5pt]
a_2 =& 0  \\[5pt]
a_3 =& \frac{1}{\beta}\left(U + a_4(1-\beta\frac{\cos\beta}{\sin\beta})\right) \\[5pt]
a_4 =& \frac{U\sin\beta\cos\beta}{\Delta}\left(\sin^2\beta - \delta^2 + \lambda(\delta\beta - \sin^2\beta)- \lambda\pi \tan\beta \right) 
\end{align*}
where $\Delta= (\sin\beta\cos\beta-\beta)(\delta^2- \sin^2\beta) + \lambda(\beta^2-\sin^2\beta)(\delta - \sin\beta\cos\beta)$ and $\delta = \beta - \pi$.
\\

Streamfunction expression for a fixed wedge is obtained by solving the biharmonic equation with the standard no-slip boundary condition over the moving wall.
The resulting expression for the streamfunction is as follows:
%
\begin{equation}\label{eq:streamfunction_HS71}
   \psi(r,\theta;\phi) = rK(\theta,\phi) = rU\left(\frac{ \theta\sin\phi \cos(\theta -\phi)-\phi\sin\theta}{\sin\phi \cos\phi-\phi}\right).
\end{equation}
%
In all our experiments, phase A is air, which has very low viscosity. Therefore, the equn. \ref{eq:streamfunction_HS71} is formulated by considering the air or gas above the liquid as passive, i.e., $\lambda = 0$. The interfacial speed, $v_i^{HS}$, is determined by setting the angle $\theta$ as $\theta = \phi$ and it is given by: %
\begin{equation}
    v_i^{HS} = v_r(r,\phi) = U \left(\frac{\sin \phi - \phi \cos \phi}{\sin \phi \cos \phi - \phi}\right).
    \label{eq:Scriven2}
\end{equation}
%
Here, $U$ is the plate speed, and $\phi$ is the contact angle for the fixed wedge. Note that the expression for interfacial speed is independent of radial position $r$. Hence, the interfacial speed for the fixed wedge is constant along the interface.  

\section{The expressions for velocity components $v_r$ and $v_{\theta}$ for curved wedge geometry} \label{appx:MWS}
The fixed contact angle $\phi$ in the streamfucntion expression (see equn.\ref{eq:streamfunction_HS71}) is replaced with interface angle $\beta(r)$ to incorporate the curvature effects. The velocity components $v_r$ and $v_{\theta}$ can be expressed as: 
\begin{eqnarray}
   && v_r(r,\theta;\beta) = \frac{\partial K}{\partial\theta}, \\
   && v_{\theta}(r,\theta;\beta) = -K - r\frac{\partial K}{\partial \beta}\frac{\partial \beta}{\partial r},
\end{eqnarray}
Using the expression for $K(\theta,\beta)$ from equn. \ref{eq:Modulate_wedge_1}, we have
\begin{align} 
\frac{v_r(r,\theta;\beta)}{U} = &\frac{\sin \beta \cos (\theta-\beta)-\beta \cos\theta - \theta \sin \beta \sin (\theta - \beta)}{ \sin\beta \cos\beta-\beta }, \label{eq:vr_MW}\\ 
\frac{v_{\theta}(r,\theta;\beta)}{U} = &\frac{\beta \sin\theta - \theta\sin \beta \cos (\theta-\beta)}{\sin\beta \cos\beta-\beta} \nonumber
   \\ &  + r\frac{d\beta}{dr} \left(\frac{2\sin^2\beta(\beta \sin\theta - \theta\sin\beta \cos(\theta-\beta))}{(\sin\beta \cos\beta-\beta)^2} - \frac{\theta \cos(2\beta - \theta)-\sin\theta}{\sin\beta \cos\beta-\beta}\right). \label{eq:vtheta_MW}
\end{align}


\section{Interface shape from GLM using dynamic contact angle} \label{appx:GLM_dynamic}
\begin{figure}[h]
\centering
\includegraphics[trim = 0mm 0mm 0mm 0mm, clip, angle=0,width=0.6\textwidth]{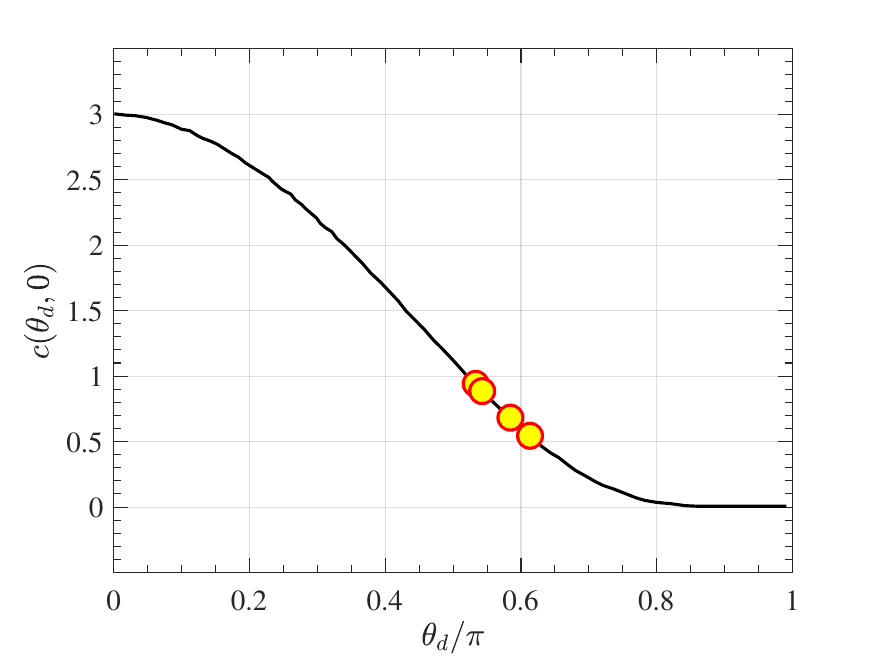}
\vspace{2mm}
\caption{Same as figure \ref{fig:c_vs_theta} with only difference in contact angle employed at the contact line. Dynamic contact angle ($\theta_d$) is used instead of $\theta_e$. Here, the filled circles denote our data points where interface shape are compared with GLM as shown in figure \ref{fig:interface_shape_GLM_thetaD}.}
\label{fig:c_vs_thetaD}
\end{figure}
\begin{figure}
\centering
\subfigure[]{
\label{fig:comparison_3mm_speed_Si_500_GLM_thetaD}
\includegraphics[trim = 0mm 0mm 0mm 0mm, clip, angle=0,width=0.48\textwidth]{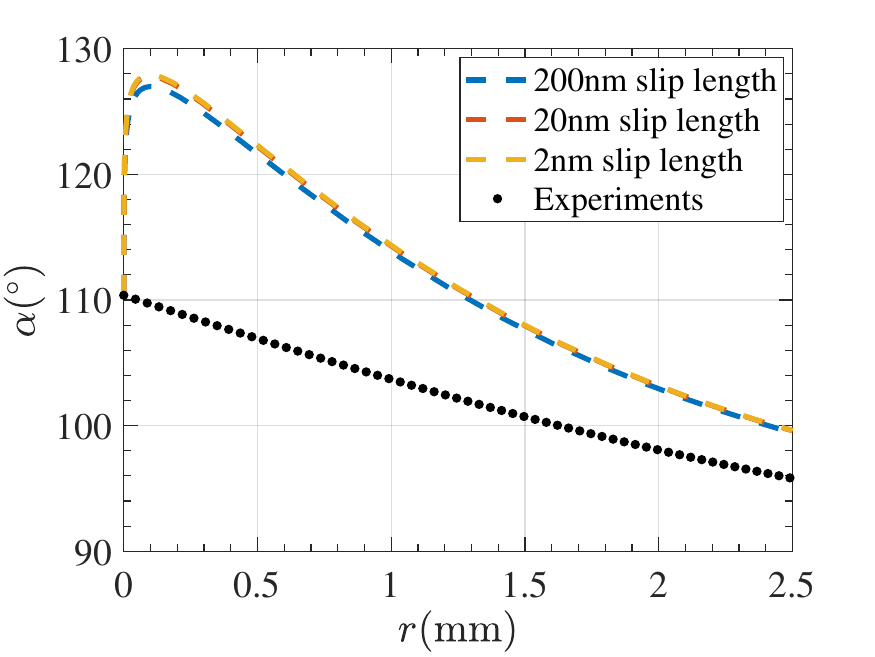}}
\subfigure[]{
\label{fig:comparison_150microns_speed_SW60_coated_GLM_thetaD}
\includegraphics[trim = 0mm 0mm 0mm 0mm, clip, angle=0,width=0.48\textwidth]{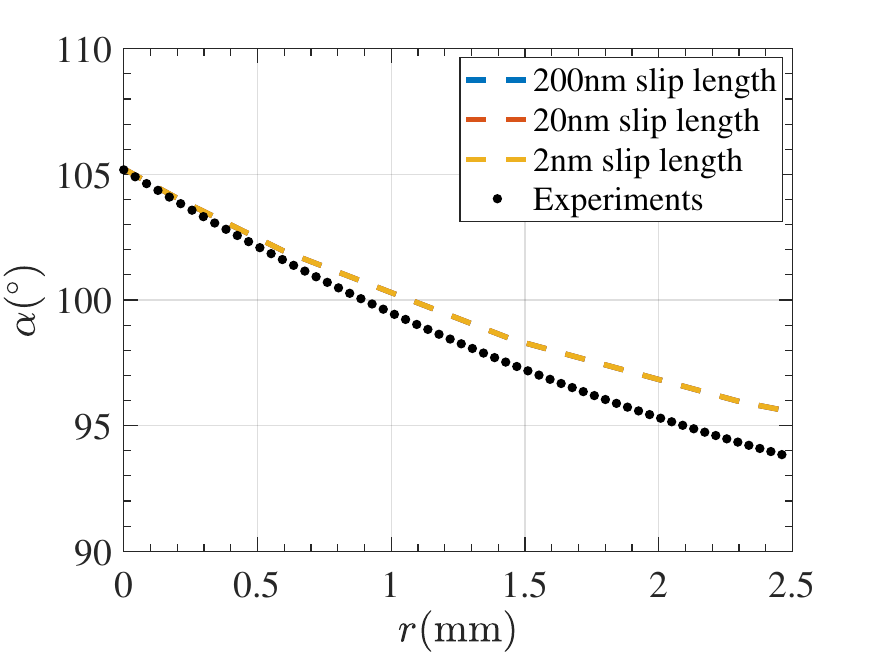}}
\subfigure[]{
\label{fig:comparison_GLM_sugar48_coated_thetaD}
\includegraphics[trim = 0mm 0mm 0mm 0mm, clip, angle=0,width=0.48\textwidth]{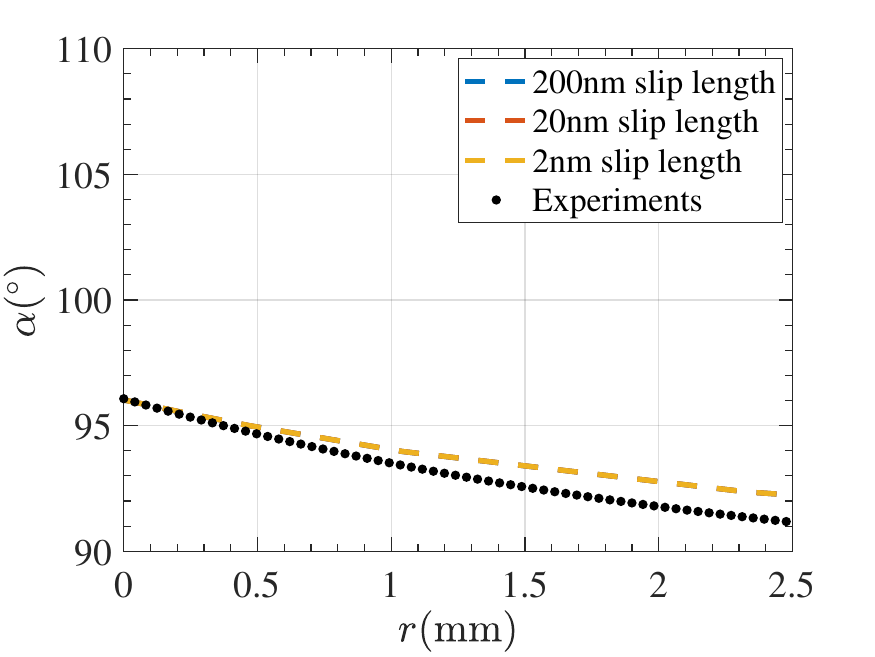}
}
\subfigure[]{
\label{fig:comparison_GLM_water_coated_thetaD}
\includegraphics[trim = 0mm 0mm 0mm 0mm, clip, angle=0,width=0.48\textwidth]{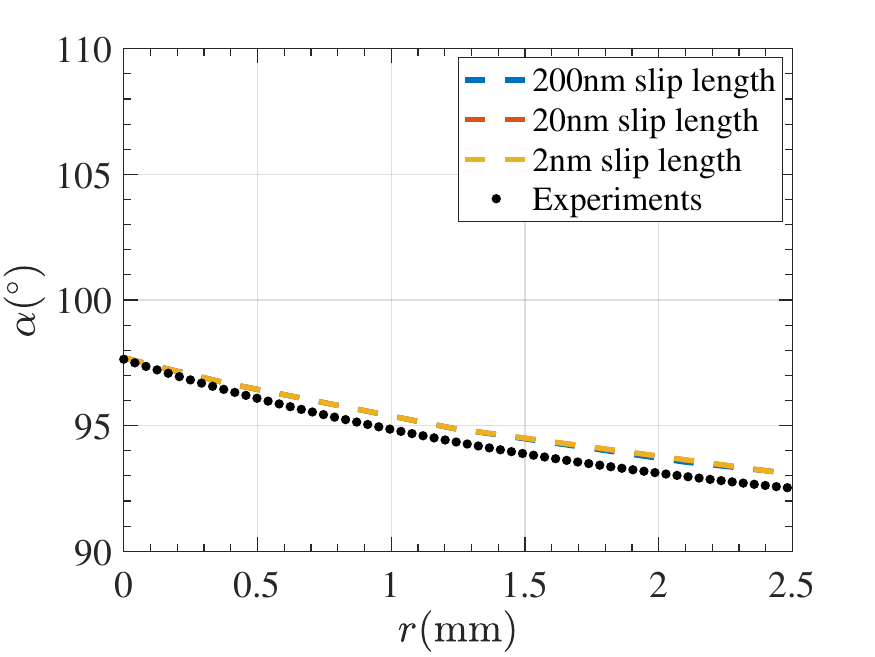}
}
\caption{Same as figure \ref{fig:interface_shape_GLM} with only difference in contact angle employed at the contact line. Dynamic contact angle ($\theta_d$) is used instead of $\theta_e$. (a) air-500 cSt silicone oil at $Ca = 8.37\times 10^{-2}$, $\theta_{d} = 110.4$; $c=0.54$; (b) air-sugar 60\% at $Ca = 1.04 \times 10^{-4}$, $\theta_{d} = 105.2$; $c=0.68$ over a hydrophobic coated solid surface. (c) air-sugar 48\% at $Ca = 8.16\times 10^{-5}$, $\theta_{d} = 96$; $c=0.94$ over a hydrophobic coated solid surface. (d) air-water at $Ca = 1.24 \times 10^{-5}$, $\theta_{d} = 97.7$; $c=0.88$ over a hydrophobic coated solid surface.}
\label{fig:interface_shape_GLM_thetaD}
\end{figure}

In \S\ref{sec:interface_shape}, the generalized lubrication model was compared with experimental data using the static advancing contact angle. Using the static
advancing angle was justified on the grounds that in the vicinity of the contact line where bulk motions cease to exist, the relevant contact angle boundary condition would be the microscopic or static (advancing) contact angle. In the present section, we present the comparison of experimental data with GLM solution, but with the dynamic contact angle. These GLM solutions are obtained by solving equations \eqref{eq:glm} - \eqref{eq:glmbc}, but with $\theta_e$ replaced with $\theta_d$ in equn. \ref{eq:glmbc}. The constant $c$ also changes for the different cases and is shown in figure \ref{fig:c_vs_thetaD} and the comparison of interface shapes is shown in figure \ref{fig:interface_shape_GLM_thetaD}. The main difference between figure \ref{fig:interface_shape_GLM} and \ref{fig:interface_shape_GLM_thetaD} can be seen in the vicinity of the contact line, i.e. as $r\rightarrow 0$. The large difference between `equilibrium' and dynamic contact angles between figures \ref{fig:comparison_3mm_speed_Si_500_GLM} and \ref{fig:comparison_3mm_speed_Si_500_GLM_thetaD} produces significant difference in the values of the constant $c$, 2.95 and 0.54 respectively. This leads to large variation in the interface shapes from GLM for the two cases. Further, for figure \ref{fig:comparison_3mm_speed_Si_500_GLM_thetaD}, the GLM solution over predicts the dynamic contact angle and a sharp 'boundary' layer is produced as $r\rightarrow 0$. Less deviation is observed for figures \ref{fig:comparison_150microns_speed_SW60_coated_GLM_thetaD}, \ref{fig:comparison_GLM_sugar48_coated_thetaD} and \ref{fig:comparison_GLM_water_coated_thetaD} from its counterpart in figure \ref{fig:interface_shape_GLM}.

\section{Streamfunction calculation from PIV data}\label{appx:streamfunction_calculation}
\begin{figure}
\centering
\subfigure[]{
   \includegraphics[trim = 0mm 0mm 0mm 0mm, clip, angle=0,width=0.5\textwidth]{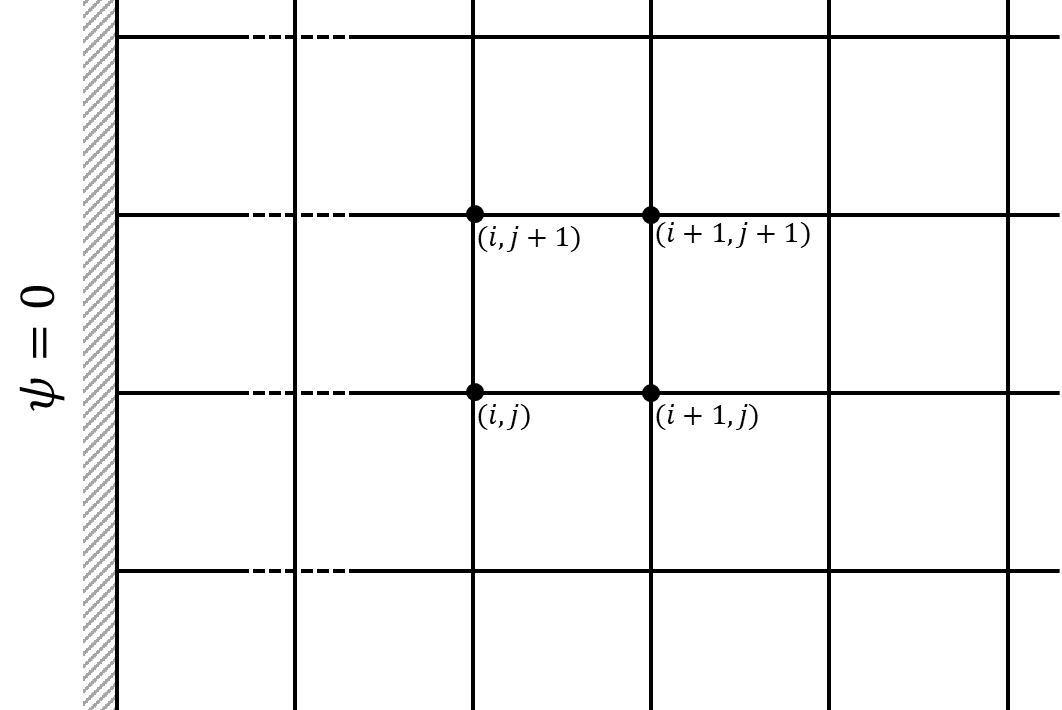}
   \label{fig:streamfunction_generator}
   }
   \hspace{12mm}
\subfigure[]{
   \includegraphics[trim = 0mm 0mm 0mm 0mm, clip, angle=0,width=0.31\textwidth]{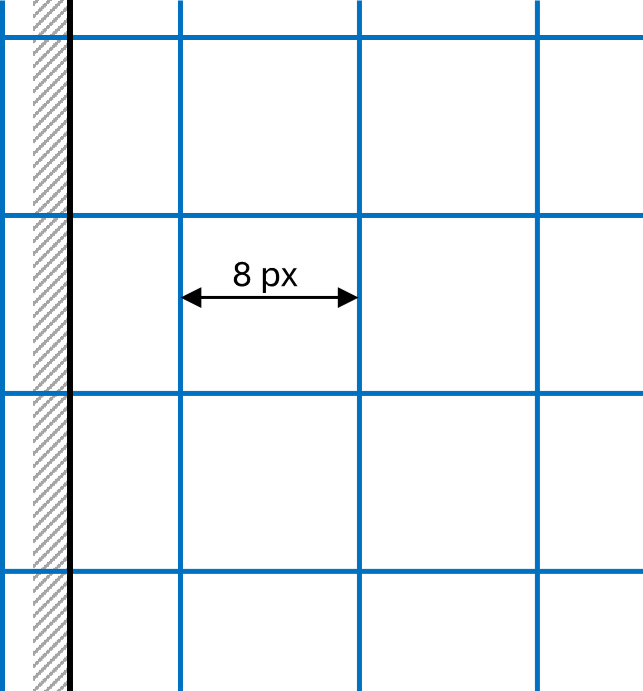}
   \label{fig:physical_space}
   }
\caption{(a) Schematic representation of the PIV grid to calculate streamfunction, (b) A schematic view of the interrogation window coincident with the physical space containing the moving plate. The typical interrogation window size used in the experiments is 16 px $\times$ 16 px with an offset of 8 px $\times$ 8 px. This provides the upper bound for the location of the first grid point where a velocity vector is obtained.}
\end{figure}

As is well known, PIV data produces a vector image with velocity field laid on a uniform grid of points. If $u(x,y)$ and $v(x,y)$ represent the horizontal and vertical components of a steady-state velocity field obtained from experiments, it is possible to determine the streamfunction $\psi(x,y)$ from this velocity field by using mass conservation. If $\dot{q}$ is the mass flux between any two points, say, 1 and 2, then it is well known that $\psi_1-\psi_2 = \dot{q}$. Using this principle, we calculate the mass flux between a pair of  grid points sequentially and determine the function $\psi(x,y)$ in the entire flow domain. This is illustrated schematically in figure \ref{fig:streamfunction_generator} and algorithmically written as
\begin{equation}
    \psi_{(i+1,j)} = \psi_{(i,j)} + \frac{v_{(i,j)}+v_{(i+1,j)}}{2}\Delta x
\end{equation}
where $\Delta_x$ is the horizontal distance between $(i,j)$ and $(i+1,j)$. Having determine $\psi_{(i+1,j)}$, we can then determine $\psi_{(i+1,j+1)}$ as
\begin{equation}
    \psi_{(i+1,j+1)} = \psi_{(i+1,j)} + \frac{u_{(i+1,j)}+u_{(i+1,j+1)}}{2}\Delta y.
\end{equation}
This process is sequentially followed over the entire domain. Since the vertical wall, represented by $i=1$, represents a streamline, the streamfunction value is constant along the wall. We set $\psi_{1,j} \equiv 0$ consistent with the boundary condition for streamfunction in theoretical models. Note that the first grid point, i.e. $i=1$, is not at the wall, but at the offset location shown in figure \ref{fig:physical_space}. The maximum error this causes is about 8 pixels, roughly about 32 $\mu$m. This means that theoretical and experimental streamfunction contours could be displaced from each other horizontally by a maximum of 32$\mu$m. At the scale of the flow field which is approximately $\tilde O(mm)$ or $\tilde O(1000\mu m)$, this error is deemed to be within acceptable limits.

\bibliography{main}

\end{document}